\documentclass[prd,aps,twocolumn,nofootinbib,showpacs,superscriptaddress]{revtex4-1}
\usepackage{amsfonts}
\usepackage{amsmath}
\usepackage{amssymb}
\usepackage{bm}
\usepackage{dcolumn}
\usepackage[dvips]{graphicx}
\usepackage{graphics}
\usepackage[latin1]{inputenc}
\usepackage{latexsym}
\usepackage{rotating}
\usepackage[colorlinks=true]{hyperref}
\usepackage{xspace} % Sensible space treatment at end of simple macros
\usepackage[usenames]{color}
\usepackage{mathrsfs}
\usepackage{multirow}
\usepackage{pifont}
\usepackage{enumitem}
\usepackage{cleveref}
\definecolor {darkgreen}{rgb}{0.2,0.7,0.2}
\definecolor{purple}{rgb}{0.5,0,0.5}

\newcommand\be{\begin{equation}}
\newcommand\ba{\begin{eqnarray}}
\newcommand\ee{\end{equation}}
\newcommand\ea{\end{eqnarray}}

\def\t{\tau}

\def\L{\Lambda}

\newcommand\bw{\begin{widetext}}
\newcommand\ew{\end{widetext}}

\newcommand{\nn}{\nonumber}

\newcommand{\PPN}{{\mbox{\tiny PPN}}}
\newcommand{\source}{{\mbox{\tiny source}}}
\newcommand{\massless}{{\mbox{\tiny massless}}}
\newcommand{\massive}{{\mbox{\tiny massive}}}

\newcommand{\SSs}{{\mbox{\tiny SS}}}

\begin{document}

\title{Cosmological Evolution and Solar System Consistency \\ of Massive Scalar-Tensor Gravity}

\author{Thibaut Arnoulx de Pirey Saint Alby}
\affiliation{eXtreme Gravity Institute, Department of Physics, Montana State University, Bozeman, MT 59717, USA.}
\affiliation{Department of Physics, \'Ecole Normale Sup\'erieure, \\ 24 rue Lhomond, 75231 Paris cedex 05, France}
\author{Nicol\'as Yunes}
\affiliation{eXtreme Gravity Institute, Department of Physics, Montana State University, Bozeman, MT 59717, USA.}

\date{\today}

%%%%%%%%%%%%%%%%%%%%%%%%%%%%%%%%%%%%%%%%%%%%%%%%%
\begin{abstract} 

The scalar-tensor theory of Damour and Esposito-Far\`ese recently gained some renewed interest because of its ability to suppress modifications to General Relativity in the weak field, while introducing large corrections in the strong field of compact objects through a process called scalarization. 
A large sector of this theory that allows for scalarization, however, has been shown to be in conflict with Solar System observations when accounting for the cosmological evolution of the scalar field.
We here study an extension of this theory by endowing the scalar field with a mass to determine whether this allows the theory to pass Solar System constraints upon cosmological evolution for a larger sector of coupling parameter space.
We show that the cosmological scalar field goes first through a \emph{quiescent phase}, similar to the behavior of a massless field, but then it enters an \emph{oscillatory phase}, with an amplitude (and frequency) that decays (and grows) exponentially. 
We further show that after the field enters the oscillatory phase, its effective energy density and pressure are approximately those of dust, as expected from previous cosmological studies.  
Due to these oscillations, we show that the scalar field cannot be treated as static today on astrophysical scales, and so we use time-dependent perturbation theory to compute the scalar-field-induced modifications to Solar System observables. 
We find that these modifications are suppressed when the mass of the scalar field and the coupling parameter of the theory are in a wide range, allowing the theory to pass Solar System constraints, while in principle possibly still allowing for scalarization.   

\end{abstract}

% 04.50.Kd Modified theories of gravity
% 98.80.-k Cosmology
% 04.25.Nx Post-Newtonian approximation; perturbation theory; related approximations
% 97.60.Jd Neutron stars

\maketitle
%\tableofcontents

\allowdisplaybreaks[4]

%%%%%%%%%%%%%%%%%%%%%%%%%%%%%%%%%%%
\section{Introduction}  
 
%Why testing GR is important.
A plethora of modified gravity theories have been constructed over the years in an attempt to provide new perspectives on fundamental physics. Although useful from a phenomenological standpoint, every time a new theory is proposed, observations of different types have shown the predictions of the theory to be in conflict with Nature. Solar System observations allow for tests of the classic predictions of General Relativity (GR)~\cite{lrr-2014-4} through, for example, the tracking of spacecrafts and the laser ranging of the Moon~\cite{lrr-2010-7}. Binary pulsars observations have also proven to be an excellent tool to test Einstein's theory in \emph{strong gravity} environments~\cite{lrr-2003-5}, i.e.~where the gravitational field is strong but not rapidly varying. The recent gravitational wave detection of the coalescence of a black hole binary~\cite{Abbott:2016blz} has taken us one step further by allowing for tests in \emph{extreme gravity}~\cite{TheLIGOScientific:2016src,Yunes:2016jcc}, i.e.~where the field is not just strong and non-linear but also highly dynamical~\cite{Yunes:2013dva}.

%Introduction to ST theories
Scalar-tensor theories of gravitation are a great example of such proposed modifications to gravity. These theories were originally introduced in the 1950's by Jordan, Brans, Dicke and Fierz~\cite{Brans:2005ra} to account for possible variations of Newton's gravitational constant $G$. These variations are typically achieved by adding one (or several) dynamical scalar fields that couple to gravity directly, but to matter indirectly. A dynamical scalar field of this type modifies gravitational physics in the Solar System, and thus, it is in conflict with experiments unless the coupling of the field to gravity is suppressed. An example of this is the Cassini Probe observation of the Shapiro time delay~\cite{Abbate:2003stu,Bertotti:2003rm}: signals from the Cassini spacecraft were observed to be (Shapiro) time delayed when the Sun was between the spacecraft (on its way to Jupiter) and Earth by exactly the amount predicted in GR~\cite{lrr-2014-4}. Scalar-tensor theories predict a correction to this effect, whose absence in the Cassini observation forces the coupling between the scalar field and gravity to be smaller than one part in $10^{5}$, limiting the interest in these theories.  

%Revival of ST theories
This interest, however, was reignited in the early 1990's, when Damour and Esposito-Far\`ese (DEF) proposed a massless scalar-tensor theory~\cite{DEF_model} with a remarkable feature:~a non-linear process could force the scalar field to induce order unity deviations in the strong field, while allowing the theory to reduce to GR in the weak field, thus avoiding Solar System constraints. This \emph{scalarization} process typically activates the scalar field when the energy of the system exceeds a certain threshold. When considering an isolated neutron star, its energy is proportional to its compactness and the process is called \emph{spontaneous scalarization}~\cite{DEF_model,Damour:1996ke}. When considering a neutron star binary, its energy is proportional to the system's gravitational potential and the process is called \emph{dynamical scalarization}~\cite{Dynamical}. When in a binary, a neutron star can also become scalarized when in the presence of an external scalar field (e.g.~produced by its companion) and the process is called \emph{induced scalarization}~\cite{Dynamical}. More recently, the interest in scalar-tensor theories was also revived because they emerge as the low energy limit of higher dimensional theories, such as string theory~\cite{Polchinski:1998rq} and Kaluza-Klein type theories~\cite{Duff:1994tn}. 
 
%Killing ST theories again.
A few years ago, however, even this DEF scalar-tensor model was shown explicitly to be in conflict with Solar System observations when accounting for the cosmological evolution of the field~\cite{Sampson:2014qqa,Anderson}. The DEF model is in agreement with observations only if the asymptotic value of the scalar field at spatial infinity is \emph{chosen} to correspond to the minimum of an effective potential that accounts for the coupling of the field to matter. The asymptotic value of the scalar field, however, cannot be freely chosen; rather, it must be consistently determined by the cosmological evolution of the scalar field. Following the work of~\citep{Damour_cosmo}, Refs.~\cite{Sampson:2014qqa,Anderson} showed that the negative-$\beta$ branch of DEF theories, i.e.~the typical branch that predicts scalarization, is exactly the one in which the field has a run-away cosmological evolution, leading to an asymptotic field value that results in grave disagreement with Solar System observations. Scalarization is still possible in the positive-$\beta$ branch, for only for a restricted class of equations of state, i.e., those for which the trace of the stress energy tensor becomes negative somewhere inside the star~\cite{Mendes:2014vna,Mendes:2014ufa,Palenzuela:2015ima,Mendes:2016fby}. 

%What we do in this paper.
This paper studies a generalized version of the DEF model that endows the scalar field with a mass in the hopes of avoiding Solar System constraints for all $\beta$ upon accounting for its consistent cosmological evolution. Indeed, establishing the consistency of massive DEF theory in both cosmological and Solar System contexts represents a interesting starting point for an extensive study of the predictions of this model in extreme gravity environments, such as in the exterior of neutron stars. Massive DEF theories are interesting because they have the potential to continue to allow for scalarization in neutron star system, provided the Compton wavelength of the field is within a certain range~\cite{Pretorius}. We here focus on the cosmological evolution of the massive scalar field and its impact on Solar System observables through numerical simulations and analytical perturbation theory; we leave a detailed study of scalarization in massive DEF theories to future work. 

%Result 1: 
We find that the cosmological evolution of the massive scalar field is dramatically different to that of the massless field. In the massless case, the scalar field presents mostly power-law behavior, with the exponents dependent on the matter-energy content of the universe. In the massive case, however, we prove that after a \emph{quiescent phase}, well-characterized by power-law behavior, the scalar field enters an \emph{oscillatory phase}. When this happens, the amplitude of the envelope of the field decays exponentially with time but the frequency of its oscillations grows exponentially, until it reaches a limiting value. 

%Consequence of Result 1
This behavior corresponds effectively to that of a dust cosmological component. We show explicitly that the effective equation of state of the scalar field averages to zero in time, and its effective cosmological energy decays inversely with the cube of the scale factor. Therefore, massive DEF is consistent with small redshift cosmological observations, the contribution of the scalar field masquerading in the $\Omega_M$ cosmological parameter, as measured for example in Type Ia supernovae~\cite{Riess:2004nr}. Indeed, minimally-coupled massive scalar tensor theories, i.e., those without a direct coupling between the scalar field and the Ricci scalar as proposed by DEF, had already been shown to present this behavior in the cosmological context (e.g., scalar field dark matter models~\cite{SFDM}).
 
%Result 2: 
Adding a mass term in the Lagrangian provides a mechanism for the scalar field to decay and its envelope to become small today relative to its initial value (i.e., the one at the end of inflation). This, however,  comes at the cost of forcing the field to also oscillate extremely rapidly today. These results are supported by both fully numerical simulations, as well as analytical calculations that characterize stationary points of the field equations through an effective Hamiltonian. We have further implemented a multiple-scale analysis to obtain an approximate solution for the evolution of the scalar field that approximates the numerical simulations very accurately. 

The time-dependence induced in the scalar field by its cosmological evolution imposes a \emph{time-dependent boundary condition at spatial infinity} when dealing with its excitations on astrophysical scales at small redshift. Consequently, time-independent perturbation theory \emph{cannot} be formally employed to study the observable consequences of this theory in the Solar System or with binary pulsars. We use time-dependent perturbation theory to study the weak field limit of the theory and find that the field around a massive astrophysical body does not present a spatial Yukawa-exponential damping with distance to the source, as one may expect in theories with massive scalars\footnote{This expectation is based on the assumption that the asymptotic value of the field is static, which is not the case in massive DEF theory for a wide range of scalar field masses.}. Instead, the cosmological evolution of the field forces it to appear massless in the Solar System (on a spacelike hypersurface) due to a cancelling mechanism between the mass term and the second time derivative of the field in its evolution equation. Consequently, Solar System constraints can still be cast in a form that is functionally analogous to that obtained in massless DEF theory, and thus, the theory passes Solar System constraints for a wide range of masses and coupling parameter values. 

%Paper organization
The remainder of this paper deals with the details of the calculations described above. Section~\ref{sec:ABC} summarizes the basics of scalar-tensor theories of gravity and provides a summary of the best current constraints on its coupling parameters. Section~\ref{sec:cosmo} presents the cosmological study of massive DEF theory and shows how this theory embeds consistently in the late time evolution of the universe. Finally, Sec.~\ref{sec:sol} provides insight into the weak field limit of the theory and its link with cosmology. Henceforth, we use units in which $c=1$ and follow the conventions of~\cite{Misner:1974qy}. For example, Greek letters in indices run over the four spacetime coordinates and Latin letters in indices represent spatial quantities.

%%%%%%%%%%%%%%%%%%%%%%%%%%%%%%%%%%%
\section{The ABC of Scalar-Tensor Theories} 

\label{sec:ABC}

This section begins by recalling the action and the field equations of scalar-tensor theories in the Jordan and Einstein frames. We then focus on DEF theory and present the current constraints on its parameters.

%------------------------------------------------------------------------------------
\subsection{Equations of Motion}  

We study scalar-tensor theories of gravity that can be described by the following action in the Jordan frame:
\begin{align}
S_{ST}[\widetilde{g}_{\mu\nu},\phi ,\psi] & = \frac{1}{2 \kappa}\int d^{4}x \sqrt{-\widetilde{g}}\left[\phi \widetilde{R} + L_{\phi}(\phi,\widetilde{g}_{\mu\nu})\right] \nn \\ & + S_{m}[\widetilde{g}_{\mu\nu},\Psi] \, ,
\end{align}
where $\kappa = 8 \pi G$ and $\widetilde{g}_{\mu\nu}$ is the Jordan frame metric. $\widetilde{R}$ and $\widetilde{g}$ are the associated Ricci scalar and determinant respectively.
The field $\phi$ is an additional scalar degree of freedom that couples non-minimally to the gravitational sector, and $\Psi$ denotes all matter degrees of freedom. This action is ``natural'' because the scalar field was originally introduced to account for possible variations of the gravitational constant. Moreover, the Lagrangian of the scalar field is written in a generic way as follows:
\begin{align}
L_{\phi}(\phi,\widetilde{g}_{\mu\nu}) = -\frac{\omega(\phi)}{\phi}\widetilde{g}^{\mu\nu} \partial_{\mu}\phi \partial_{\nu}\phi - \Pi(\phi)\,,
\end{align}
i.e.~as the sum of a kinetic term and a potential.  

The absence of a direct coupling between $\phi$ and $\Psi$ guarantees that the weak equivalence principle is not violated. This principle states that the motion of a freely falling test mass is independent of its internal structure and composition, and it has been experimentally well-verified. Even in the absence of a $\phi$-$\Psi$ coupling, however, the addition of a new scalar degree of freedom does violate the strong equivalence principle. That is, the motion of a freely-falling, \emph{self-gravitating} body does depend on its internal structure, because the latter affects the scalar field, and this field contributes to the motion.

In the Jordan frame, variation of the action with respect to $\widetilde{g}_{\mu\nu}$ and $\phi$ yields the following field equations:
\begin{align}
\phi \left(\widetilde{R}_{\mu\nu} - \frac{1}{2}\widetilde{R} \widetilde{g}_{\mu\nu}\right)& =  \, \kappa \widetilde{T}_{\mu\nu}^m + \left(\nabla_{\mu}\nabla_{\nu}\phi - \widetilde{g}_{\mu\nu}\nabla_{\alpha} \nabla^{\alpha}\phi\right) \nn \\
 & +  \frac{\omega (\phi)}{\phi}\left(\partial_{\mu}\phi \partial_{\nu}\phi - \frac{1}{2}\widetilde{g}_{\mu\nu}\partial^{\alpha}\phi \partial_{\alpha}\phi\right) \nn \\
 & - \widetilde{g}_{\mu\nu}\frac{\Pi(\phi)}{2 \phi}\,,  \\
[2\omega(\phi) + 3]\nabla_{\alpha} \nabla^{\alpha}\phi & =  \, \kappa \widetilde{T}^m - \frac{d\omega}{d\phi}\partial^{\mu}\phi \partial_{\mu}\phi + \frac{d\Pi}{d\phi}\phi \nn \\
& - 2 \Pi(\phi)\,,
\end{align}
where 
\begin{align}
\widetilde{T}_{\mu\nu}^m = \frac{2}{\sqrt{-\widetilde{g}}} \frac{\delta S_{m}[\widetilde{g}_{\mu\nu},\Psi]}{\delta \widetilde{g}_{\mu\nu}} \, ,
\end{align}
is the usual matter stress energy tensor and $T^m$ its trace. Because ordinary matter only couples to the metric, the Jordan frame stress energy tensor is covariantly conserved, as required by the weak equivalence principle.

In order to simplify the mathematics, the action can be rewritten in the \emph{Einstein frame}, characterized by the metric $g_{\mu\nu}$, which is conformally related to $\widetilde{g}_{\mu\nu}$ by $\widetilde{g}_{\mu\nu} = A(\varphi)^2 g_{\mu\nu}$. Here, $\varphi$ is a new scalar that satisfies 
\begin{align}
A(\varphi)^2  & = \, \phi^{-1}\,, \\
\left(\frac{d \ln A}{d \varphi}\right)^2 & = \, [2 \omega(\phi)+ 3]^{-1}\,.
\end{align} 
The Einstein frame action is then
\begin{align}
\label{actionEin}
S_E & =  \int d^{4}x \frac{\sqrt{-g}}{2\kappa}\left[R - 2 g^{\mu\nu}\partial_{\mu}\varphi \partial_{\nu}\varphi - 2 V(\varphi)\right] \nn \\
& + S_{m}[A^{2}(\varphi )g_{\mu\nu},\Psi]\,,
\end{align}
where the $\varphi$ potential is related to the $\phi$ potential via 
\begin{align}
V(\varphi)=\frac{1}{2}\frac{\Pi(\phi)}{\phi^2}\,.
\end{align}
In this frame, the field equations are
\begin{align}
\label{eq_ein}
R_{\mu\nu} - \frac{1}{2}R g_{\mu\nu} & = \, \kappa T^{m}_{\mu\nu} + T^{\varphi}_{\mu\nu}\, , \\
\label{eq_KG}
g^{\mu\nu} \nabla_{\mu} \nabla_{\nu} \varphi & =  \, -\frac{\kappa}{2}\alpha(\varphi)T^{m} + \frac{1}{2}\frac{dV}{d\varphi} \, ,
\end{align}
where we have defined
\begin{align}
T^{\varphi}_{\mu\nu} &= 2 \partial_{\mu}\varphi \partial_{\nu}\varphi - g_{\mu\nu}\partial^{\alpha}\varphi \partial_{\alpha}\varphi  - g_{\mu\nu}V(\varphi)\, ,
\end{align}
and
\begin{align}
\label{ST}
T_{\mu\nu}^{m} &= \frac{2}{\sqrt{-g}} \frac{\delta S_{m}[A^{2}(\varphi )g_{\mu\nu}, \Psi]}{\delta g_{\mu\nu}} \, ,
\end{align}
is the Einstein frame stress-energy tensor, with $T^{m}$ its trace with respect to the $g$ metric. The quantity
$\alpha(\varphi)={d \ln A}/{d \varphi}$ plays the role of an effective coupling function between ordinary matter and the scalar field. In addition to the mathematical simplicity of the equations, the Einstein frame has another more fundamental advantage: the spin 2 and the spin 0 degrees of freedom linearly decouple, whereas they are mixed in the Jordan frame.

%------------------------------------------------------------------------------------
\subsection{DEF Theory}

%-----------------
\subsubsection{Formulation} 
DEF theory~\cite{DEF_model}  is most easily formulated in the Einstein frame and it corresponds to a first order development in the field of the effective coupling function $\alpha(\varphi)$, such that
\begin{align}
A(\varphi) = e^{\gamma \varphi + \frac{1}{2}\beta \varphi^2}\,,
\end{align}
where $\gamma$ and $\beta$ are two, free coupling parameters of the theory. In the original Brans-Dicke theory ($\beta = 0$), $\gamma$ is related to the more common $\omega_{BD}$ parameter via
\begin{align}
\gamma^2 = \,(2 \omega_{BD}+ 3)^{-1}\,.
\end{align}

The choice of the potential determines whether the field is massive or not. In the standard DEF model, $V(\varphi) = 0$, and the field is massless. In this paper, however, we allow the field to have a mass by setting
\be
V(\varphi) = m^2\varphi^2\,,
\ee
where $m$ is the mass of $\varphi$. This means that in the Jordan frame, the field $\phi$ evolves in a potential of the form
\be
\Pi(\phi) = 2  m^{2} \; \phi^{2}  \varphi(\phi)^{2}\,,
\ee
which is not a simple quadratic $\phi$ potential. From now on, we focus only on massive DEF theory.

%-----------------
\subsubsection{Constraints}
\label{constraints} 
There exist mainly three approaches to constrain modified theories of gravity that respect the weak equivalence principle. First, the theory must reproduce GR very accurately in the Solar System. Tests such as the measurement of the Shapiro time delay by the Cassini spacecraft have greatly constrained the coupling constant measured in the Solar System of the original Brans-Dicke theory [$(\gamma,\beta,m)=(\gamma,0,0)$] and more generally in massless DEF theory [$(\gamma,\beta,m)=(\gamma,\beta,0)$]. In fact, in massless theories, the $\gamma_{\PPN}$ parameter, which measures the amount of spatial curvature created by a unit rest mass \footnote{This parameter should not be confused with the $\gamma$ coupling parameter of DEF theory.}, and which is equal to 1 in GR, is related to the coupling constant measured at spatial infinity by~\cite{will}
\begin{align}
\label{cons}
\alpha_{\massless}(\varphi_0)^2 = \frac{1-\gamma_{\PPN}}{1+\gamma_{\PPN}}\,,
\end{align}
where $\varphi_0$ is the \emph{asymptotic value of the field at spatial infinity}, or simply its cosmological value for short. The previous equation, combined with experimental data, imposes the constraint
\begin{align}
\label{cons_bis}
\alpha_{\massless}(\varphi_0)^2 \leq 10^{-5}\,.
\end{align}

The previous inequality suggests that important deviations from GR and consistency with Solar System observations are incompatible, and this discouraged the community from studying scalar-tensor theories for many years. This conclusion, however, was proven to be incorrect in general, with massless DEF theory in the $\gamma \ll 1$ and $\beta < 0$ sector as a particular counter-example. In this case, the constraint above becomes
\begin{align}
\label{eq:masslessDEF-cons}
\alpha_{\massive}(\varphi_0)^2 = \gamma + \beta \varphi_{0} \leq 10^{-5}\,,
\end{align}
which is satisfied, for example, if $\varphi_{0} \ll 1$ and $\gamma \ll 1$ for any $\beta$ that is not too large. But even if this constraint is saturated, massless DEF theory still predicts order unity deviations from GR in strongly gravitating, non-vacuum environments (typically neutron stars). This is achieved through a non-perturbative process called \emph{scalarization}~\citep{DEF_model}, which is analogous to spontaneous magnetization in ferromagnetism. In this process, the scalar field suddenly activates when the binding energy of the system exceeds a particular threshold (either in isolation or in a binary), or when the field is in the presence of another external field~\cite{Damour:1996ke}. 

When scalarization occurs in a binary system, the scalar field that activates is anchored to each neutron star, forcing the field to become dynamical as the objects orbit around each other. This motion induces a wave-like behavior in the scalar field, which then carries energy and momentum away from the binary, accelerating its rate of inspiral. This is where the second approach to constrain modified gravity theories comes in: binary pulsar observations. These observations allow us to track the orbital motion of a pulsar in a binary system extremely accurately. If these observations are done for a sufficiently long time, one can observe the rate of change of the orbital period due to the orbital energy decay. Such observations have been done to incredibly precision and the predictions of GR have been verified. This, in turn, implies that the binary pulsar observed could not have scalarized, which then imposes the constraint $\beta \geq -4.5$  in the negative-$\beta$ branch of massless DEF theory~\citep{beta_constraint}. Similar constraints can be placed in massive Brans-Dicke theory [$(\gamma,\beta,m)=(\gamma,0,m)$], as done in~\citep{Berti}.

The last route to constrain modified gravity theories is to study their cosmological evolution. In~\citep{Damour_cosmo}, Damour and Nordtvedt showed that GR is a cosmological (exponential) attractor in massless DEF theories when $\beta > 0$. References~\cite{Sampson:2014qqa,Anderson}, however, showed that when $\beta < 0$ GR is a cosmological (polynomial) repeller, i.e.~the scalar field diverges as $t \rightarrow \infty$, with $t$ being cosmological time. If this is the case, the constraint in Eq.~\eqref{eq:masslessDEF-cons} can only be passed for vanishingly small and highly fine-tuned values of $\gamma$ and $\beta$, or for a vanishingly small set of initial values of the cosmological scalar field at the beginning of the radiation-dominated era. Therefore, massless DEF theory with $\beta < 0$ is in conflict with Solar System constraints when one accounts for the cosmological evolution of the scalar field. 

Experimental constraints on DEF theory with a potential have not yet been explored deeply. For example, it has been recently shown by Pretorius and Ramazano$\breve{\text{g}}$lu in~\citep{Pretorius} that spontaneous scalarization of isolated and stationary neutron stars can still happen in massive theory for $\beta \leq 0$.  As in the massless case, this could provide a way to constrain the free parameters of the theory, but in this case the constraints would be on a 2-dimensional surface, i.e.~the $m$--$\beta$ space. This work, however, was conducted assuming a static asymptotic value of the scalar field, which as we will show here is not a valid assumption in a range of scalar field masses.

For the reasons described above, we restrict ourselves to massive DEF theories with $\gamma = 0$ and $\beta \leq 0$ for the rest of this paper, this last inequality being motivated by the study of spontaneous scalarization in neutron stars in massless DEF theory. We expect that when $\beta > 0$ scalarization will also arise in neutron stars in the massive case, as shown in the massless case by~\cite{Mendes:2014vna,Mendes:2014ufa,Palenzuela:2015ima,Mendes:2016fby}; there is less motivation, however, to add a mass to DEF theory in the positive $\beta$ branch, since, as discussed above, this branch automatically leads to a theory that passes Solar System constraints and is cosmologically viable.  

%%%%%%%%%%%%%%%%%%%%%%%%%%%%%%%%%%%
\section{Cosmology in massive DEF theory}
\label{sec:cosmo}
This section begins by adapting the Einstein frame field equations to a Friedmann-Lema\^{i}tre-Robertson-Walker (FRLW) metric~\cite{Friedmann:1924bb,Lemaitre:1933gd,Robertson:1933zz}. We then continue by numerically solving these equations in their exact form, and then interpreting them analytically through perturbation theory. We conclude this section with a description of how massive DEF theory embeds consistently into late-time cosmology.

%------------------------------------------------------------------------------------
\subsection{Field Equations}
Under the assumptions of homogeneity and isotropy, the Jordan frame metric is simply the FRLW one:
\begin{align}
d\widetilde{s}^2 = - d\widetilde{t}^2 + \widetilde{a}(\widetilde{t})^2 d\widetilde{l}^2\,,
\end{align}
where $ d\widetilde{l}^2 = {d\widetilde{r}^2} ({1- k \widetilde{r}^2})^{-1} + \widetilde{r}^2(d\theta^2 + \sin{\theta}^2 d\psi ^2) $. From now on, we consider a spatially-flat geometry, i.e.~$ k = 0$, as suggested by WMAP and Planck data~\cite{Ade:2013zuv}. Using $ds^{2} = e^{-\beta \varphi^2} d\widetilde{s}^2$, the Einstein metric becomes
\begin{align}
ds^{2} = - dt^2 + a(t)^2 dl^2 \, ,
\end{align}
with $ dt = e^{-\frac{1}{2}\beta \varphi^2} d\widetilde{t}$ and $ a(t) = e^{-\frac{1}{2}\beta \varphi^2}\widetilde{a}(\widetilde{t})$ the Einstein-frame scale factor.

Let us model the matter content of the universe as a sum of non-interacting perfect fluids. For any individual component, the Jordan frame stress-energy tensor is simply $ \widetilde{T}_{\mu\nu}^m = (\widetilde{\rho}+\widetilde{P})\widetilde{u}_{\mu}\widetilde{u}_{\nu} + \widetilde{P} \widetilde{g}_{\mu\nu} $, where $\widetilde{\rho}$ and $\widetilde{P}$ are the density and pressure of the particular fluid component in the Jordan frame. Equation \eqref{ST} implies that the corresponding Einstein frame stress-energy tensor also takes a perfect fluid form, with 
\begin{align}
\label{perfect_fluid}
\rho = & \, e^{2 \beta \varphi ^2} \widetilde{\rho} \, , \\
P = & \, e^{2 \beta \varphi ^2} \widetilde{P} \, ,
\end{align}
For cosmological matter sources, it is convenient to adopt the cosmological equation of state $\widetilde{P}=\varpi\widetilde{\rho}$, with $\varpi$ a constant that depends on the particular fluid component. 

In the Einstein frame, the modified Friedmann equations are
\begin{align}
3 H^2 = & \, \kappa \sum_i \rho_i + \dot{\varphi}^2 + V(\varphi)\,, \\
\label{secondfird}
-2\frac{\ddot{a}}{a}-\left(\frac{\dot{a}}{a}\right)^2 = & \, \kappa\sum_i \varpi_i \rho_i + \dot{\varphi}^2 - V(\varphi)\,,
\end{align}
where the overhead dot stands for time differentiation in the Einstein frame, $H = {\dot{a}}/{a}$ and where the sums run over all the components of the cosmological fluid.
The modified Klein-Gordon equation takes the following form
\begin{align}
\label{cosmo_KG}
\ddot{\varphi} + 3 H \dot{\varphi} = - \beta \frac{\kappa}{2} \varphi \sum_i (1-3\varpi_i) \rho_i - m^2 \varphi\,.
\end{align}

In order to  decouple these equations for the evolution of $H$ and $\varphi$, following \citep{Damour_cosmo},
we introduce the dimensionless time $p$ defined as $ dp = d\ln{a}$ and we choose $p=0$ today. With this time coordinate, the first Friedmann equation becomes
\begin{align}
\label{firstfrid}
3 H^2\left(1-\frac{1}{3}\varphi'^2\right) = \kappa \sum_i \rho_i + V(\varphi)\,,
\end{align}
where the prime stands for differentiation with respect to $p$. From the positivity of the energy density, this equation implies $\varphi' \leq \sqrt{3}$. Inserting Eqs.~\eqref{firstfrid} and~\eqref{secondfird} into~\eqref{cosmo_KG}, we find
\begin{align}
\label{field_tot}
\frac{2 \varphi''}{3-\varphi'^2}\left(\sum_i \rho_i + \frac{m^2}{\kappa}\varphi^2\right) + \varphi'\left[\sum_i(1-\varpi_i)\rho_i + 2\frac{m^2}{\kappa}\varphi^2\right] \nn \\ =  -\beta\varphi \sum_i (1-3\varpi_i)\rho_i - 2\frac{m^2}{\kappa}\varphi\,.
\end{align}
The only unknown here is $\varphi(p)$ because the covariant conservation of the individual Jordan-frame stress-energy tensors requires that for each non interacting fluid
\begin{align}
\widetilde{\rho_i}(a) &= \, \widetilde{\rho}_{i,0} \; \widetilde{a}^{-3(1+\varpi_i)} \, ,
\end{align}
where $\widetilde{\rho}_{i,0}$ is the energy density of the $i^{th}$ cosmological fluid as measured today.
Inserting Eq.~\eqref{perfect_fluid} and the relation between $a(t)$ and $\widetilde{a}(\widetilde{t})$ into the previous equation gives
\begin{align}
\label{energy}
\rho_i(p) &= \, \widetilde{\rho}_{i,0} \; e^{-3(1+\varpi_i)p} \; e^{\frac{1}{2} (1-3\varpi_i) \beta\varphi^2} \, ,
\end{align}

In the massless case and for a single cosmological fluid component, the evolution equation for the scalar field [Eq.~\eqref{field_tot}] becomes independent of $\rho$ and it reduces to 
\begin{align}
\frac{2 \varphi''}{3-\varphi'^2} + \varphi'(1-\varpi) =  -\beta\varphi(1-3\varpi) \, ,
\end{align}
which is analogous to the equation of motion of a relativistic damped harmonic oscillator. From this equation, it is clear that $\beta \leq 0$ forces the scalar field to diverge as $p \rightarrow +\infty$ because $(1-3\varpi) \geq 0$ for dust, radiation and dark energy. This justifies the constraint given in Sec.~\ref{constraints} and obtained formally in \citep{Anderson}. In the case of a massive scalar field, the oscillator analogy does not hold anymore -- or at least, it does not provide a direct way to understand the behavior of the field, forcing us to consider full numerical solutions to the above equations.

%------------------------------------------------------------------------------------
\subsection{Numerical Evolution}
\label{num}

In order to solve Eq.~\eqref{field_tot} numerically, we need to first specify both the content of the matter stress-energy tensor and the initial conditions at the beginning of simulation. For reasons that will become clear below, we choose to incorporate in our simulations only radiation, baryonic matter (in the form of dust) and dark energy with a present energy density equal to that measured today. The dark matter component of the universe has not been taken into account here; nevertheless, since this would behave in the same exact way as baryonic matter on cosmological scales, adding an additional cold dark matter fluid does not change the qualitative behavior of the results presented below. We also assume that the energy density of baryons scales as $a^{-3}$ -- even though inexact because of the coupling with radiation, we think this is not a source of error as baryons begin to dominate approximately at the same time as when they decouple from photons. In the following, we refer to the ``radiation- (matter- or dark energy-) dominated era'' as the period in the life of the universe when radiation (matter or dark energy) dominates the matter stress-energy tensor. Because the scalar field also has a non-vanishing energy density, this is not equivalent to saying that radiation (matter or dark energy) dominates the total stress-energy tensor. 

\begin{figure*}[htb]
\begin{center}
\includegraphics[width=\columnwidth,clip=true]{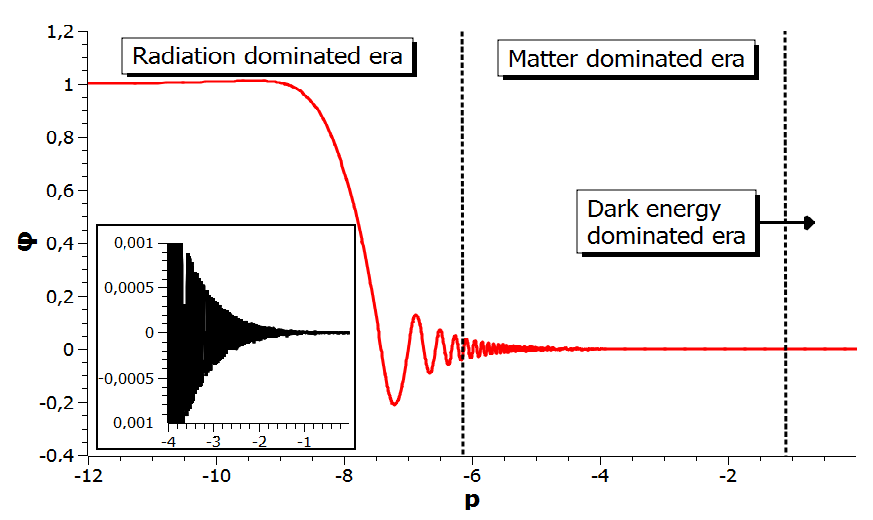}  \quad 
\includegraphics[width=\columnwidth,clip=true]{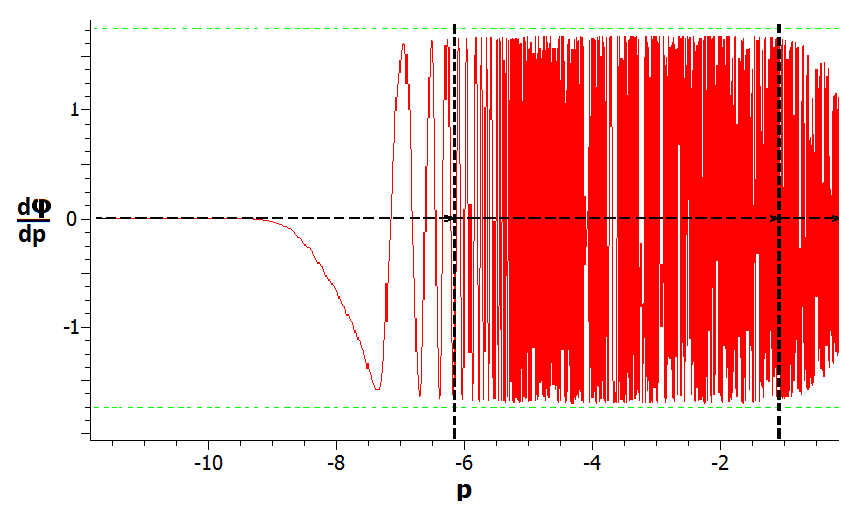}  
\caption{\label{fig:field} (Color online) Evolution of the scalar field (left panel) and its derivative (right panel) as functions of the dimensionless conformal $p$-time in massive DEF theory with $\beta = -4.5$ and $m=5.8 \times 10^{-28} \text{eV}$. In both panels, the dashed vertical lines roughly separate the different cosmological eras (radiation, matter and dark energy). Observe that after a quiescent phase, the field begins to oscillate rapidly while its amplitude decays exponentially to zero. The amplitude of the envelope of its derivative grows during radiation and matter domination, approaching the limiting value of $\sqrt{3}$ (horizontal dashed green lines on right panel) but decays during dark energy domination. The inset in the right panel shows the decay of the scalar field during the matter-dominated era and the dark-energy dominated era for $p \geq -4$.}
\end{center}
\end{figure*}

We choose to start our simulations at the beginning of the radiation-dominated era, and thus, the initial conditions are given by the behavior of the field at the end of inflation, which are not known. However, if we go backward enough in time to the very beginning of the radiation-dominated era, the field appears massless (i.e.~$ m^2 \varphi^2 \ll \kappa \rho_r$ and $m^2 \ll \kappa \rho_r$ with $\rho_r$ the energy density of radiation). In this case, and with $\varpi=1/3$, the equation of motion becomes approximately
\begin{align}
\label{massless}
\frac{2 \varphi''}{3-\varphi'^2} + \frac{2}{3} \varphi' = 0 \,.
\end{align}
As proven by Damour and Nordtvedt~\cite{Damour_cosmo}, if the velocity of the field at the end of inflation is not close to $\sqrt{3}$, then it will come to rest rapidly shortly after the beginning of the radiation-dominated era, and the magnitude of the field will vary by an amount of order unity. We thus choose for our initial conditions $(\varphi=\varphi_0,\varphi'=0)$ at the beginning of the radiation-dominated era, where $\varphi_{0}$ is a constant. In our simulations, we choose $\varphi(0)$ to be of order unity, which is a reasonable assumption, as this quantity corresponds to a Planck scale excitation of the field (see Eq.~\eqref{actionEin}) at the end of inflation. As we will show later, this assumption is consistent with constraints from Big Bang Nucleosynthesis (BBN)~\cite{2011LRR....14....2U}. 

With these initial conditions at hand, we now numerically solve the full evolution equations, presented earlier in Eq.~\eqref{field_tot}, without any approximations. Figure~\ref{fig:field} shows the numerical solution of the scalar field (left panel) and its derivative (right panel) as functions of $p$-time for $\beta = -4.5$ and $m=5.8 \times 10^{-28} \text{eV}$. During the early radiation-dominated era ($p \lesssim -9$), the field is in a \emph{quiescent phase}, presenting a quasi-stationary (approximately polynomial-like) behavior. During the late radiation-dominated era ($p \gtrsim -9$), the field enters an \emph{oscillatory phase} with the amplitude of its envelope decaying exponentially and the amplitude of the envelope of its derivative growing and approaching the limiting value of $\sqrt{3}$. This transition coincides with the time at which the radiation energy density decays enough to be of the same order as $m^2$, i.e. from Eq.~\eqref{energy}, this transition occurs when $\rho_{r,0} e^{-4p} \sim m^2/\kappa$, where $\rho_{r,0}$ is the cosmological radiation energy density today. 

From the numerical evolution, we can then approximate the behavior of the field via
\begin{equation}
\varphi(p) = A e^{-p/\tau} \cos{[\omega (p)]}.
\end{equation}
and thus,
\begin{equation}
\varphi'(p) = -\frac{A}{\tau} \cos{[\omega (p)]} - A \omega'(p) e^{-p/\tau} \sin{[\omega (p)]} \,.
\end{equation}
Clearly, the only term that is possibly responsible for the growth of the envelope of the derivative of the field is $\omega'(p) e^{-p/\tau}$, which means the frequency of the oscillations have to increase exponentially with the same characteristic time as the decay of the envelope of the field. 

As we can see from Fig.~\ref{fig:field}, the behavior remains the same during the matter-dominated era, while during the dark energy-dominated era the behavior changes. In the former, the amplitude continues to decay exponentially, as shown in the inset. In the latter, the amplitude of the derivative of the field starts to decay as well, showing that the frequency saturates (or at least does not increase as fast as the amplitude of the field decreases).

The main conclusion of our numerical simulations is that, despite the coupling to matter, which tends to induce a cosmological condensation of the field (i.e.~a divergence of the amplitude of the field on cosmological scales), the mass term seems to provide a \emph{stabilizing mechanism} that forces the field to exponentially decay, even for $\beta \leq 0$. Picking other values of $\beta$ and $m$ does not change the general behavior we discussed here. However, as the transition between steady-state and oscillations occurs when the energy density of the scalar field is of the same order as the energy density of the other cosmological fluids, this transition is delayed as one lowers $m$.

%------------------------------------------------------------------------------------
\subsection{BBN Constraints}
\label{num-BBN}

One expects that the choice of initial conditions at the beginning of the radiation-dominated era should determine whether BBN constraints are satisfied. This is because the standard scenario for the formation of light elements depends sensitively on the physical (Jordan frame) Hubble parameter $\widetilde{H}$ during BBN, which in turn depends on the initial conditions at radiation-domination. To see this explicitly, let us assume that the transition between steady state and oscillations happens after BBN, which is expected to take place during the radiation-dominated era at temperatures between $10^{-1}$ and 10 MeV. If so, neglecting $\varphi '$ and $V(\varphi)$ in Eq.~\eqref{firstfrid}, one finds $H^2 \simeq \kappa / 3 \rho_r$ so that
\begin{align}
\widetilde{H}  = & \, \, H e^{-\beta \varphi^2/2} \left( 1 + \beta \varphi \varphi '\right) \,, \\
              \simeq & \, \, H e^{-\beta \varphi^2/2}
             \simeq \, \, e^{\beta \varphi^2/2} \sqrt{\frac{\kappa}{3} \widetilde{\rho}_r} \,.
\label{eq:newH}
\end{align}
Moreover, our previous analysis showed that $\varphi$ remains constant from the beginning of the radiation-dominated era until either the beginning of the matter-dominated era or the transition to the oscillatory phase, which we have here assumed occurs after BBN. If so, the physical Hubble parameter  depends exponentially on the value of the scalar field during BBN, which is approximately the same as the initial value of the scalar field at start of radiation domination. 

Modifications to the standard scenario of light element formation are typically quantified by the speed-up factor $\xi_{bbn} = \widetilde{H}/H_{GR}$, which is constrained to satisfy $|1 - \xi_{\rm bbn}|\leq 1/8$, given current measurements of the abundance of Helium in the universe. Using Eq.~\eqref{eq:newH}, the speed-up factor becomes
\begin{equation}
\xi_{\rm bbn} \simeq  e^{\beta \varphi_{\rm init}^2 / 2} \,,
\end{equation}
where we have also used the fact that $\varphi_{\rm BBH} \sim \varphi_{\rm init}$ and that Solar System constraints force the bare coupling constant and the measured one to be approximately the same. The current abundances of Helium then impose the following constraint on the initial conditions 
\begin{equation}
\varphi_{\rm init}^2 < \frac{2}{\beta} \ln\left(\frac{7}{8} \right) \,.
\end{equation}
This justifies the relevance of the parameters we have considered in our numerical simulations ($\beta$ and $\varphi_{\rm init}$ of order unity). 

Please note that the calculation presented above is only valid if $m^2 \varphi_0^2 \ll \kappa \widetilde{\rho}_{r,0} a_{\rm bbn}^{-4}$ i.e. 
\begin{align}
m \varphi_0 \ll &  \, \, \sqrt{3 \Omega_r} H_0 \left(\frac{T_{\rm bbn}}{T_0}\right)^2 \simeq \, \, 10^{-17} \, {\rm{eV}} \,,
\end{align}
If this bound is saturated, which is far from being the case in our numerical simulations, it is more difficult to make a prediction since $\xi_{\rm bbn}$ itself will inherit the oscillating structure of the solution of the field equations. Such studies are beyond the scope of this paper. 
 
%------------------------------------------------------------------------------------
\subsection{Analytical Phenomenology}
In this section, we provide some analytical insight into the exact, numerical evolution of the scalar field presented above, using a Hamiltonian approach first and then a multiple scale analysis approach. In the following, in order to simplify the mathematics, we approximate the matter content of the universe  via a piecewise decomposition with only radiation (respectively baryonic matter or dark energy) during radiation (respectively baryonic matter or dark energy) dominated era.

%-----------------
\subsubsection{Hamiltonian Approach}
\label{subsec:H}
Let us begin by computing the Hamiltonian corresponding to the field equation [Eq.~\eqref{field_tot}], which we rewrite as
\begin{align}
\label{eq_sta}
\frac{2 \varphi''}{3-\varphi'^2} + \varphi'\frac{\left[(1-\varpi)+ 2\widetilde{m}^2\varphi^2\right]}{\left(1 + \widetilde{m}^2\varphi^2\right)} \nn \\
 =  -\left[\beta(1-3\varpi) + 2\widetilde{m}^2\right]\frac{\varphi}{\left(1 + \widetilde{m}^2\varphi^2\right)}\,,
\end{align}
where we have defined the dimensionless (but time-dependent) mass $\widetilde{m}^2 = {m^2}/({\kappa \rho})$. We identify this equation as that of a relativistic field in a time-dependent potential $V(\varphi,p)$ with a time-dependent drag force and a $\varphi'$-dependent mass term.

The Lagrangian that reproduces the conservative part (i.e., the part without the drag force) of the equation of motion is (see also \citep{Damour_cosmo})
\begin{align}
\mathcal{L}(\varphi,\varphi',p) = \left(1 + \frac{\varphi'}{\sqrt{3}}\right)\ln{\left(1 + \frac{\varphi'}{\sqrt{3}}\right)} \nn \\
 + \left(1 - \frac{\varphi'}{\sqrt{3}}\right)\ln{\left(1 - \frac{\varphi'}{\sqrt{3}}\right)} - V(\varphi,p)  \, ,
\end{align}
where $V(\varphi,p)$ can be interpreted as the potential in which the field evolves. The Euler-Lagrange equation then imposes
\begin{align}
\label{EulerLagrange}
\frac{\partial V}{\partial \varphi } = \left[\beta(1-3\varpi) + 2\widetilde{m}^2\right]\frac{\varphi}{\left(1 + \widetilde{m}^2\varphi^2\right)} \,.
\end{align}

We are mostly interesting in the late time behavior of the scalar field, which according to the exact solution presented in Fig.~\ref{fig:field} satisfies $\varphi \ll 1$, in turn allowing us to simplify the Lagrangian and its associated Hamiltonian significantly. Neglecting the $\varphi$ dependence of $\rho$ (i.e.~of $\widetilde{m}$) in the potential, 
\begin{align}
V(\varphi,p) = \left[\beta(1-3\varpi) + 2\widetilde{m}^2\right]\ln {\left( 1 +\widetilde{m}^2 \varphi^2 \right)} \frac{1}{2\widetilde{m}^2} \,.
\end{align}
The associated Hamiltonian is then found to be
\begin{align}
\label{eq:Hamiltonian}
\mathcal{H}(p,\varphi,\varphi') & = \varphi' \frac{\partial \mathcal{L}}{\partial \varphi'} - \mathcal{L} \nn \\
                                                        & = - \ln{\left(1 - \frac{\varphi'^2}{3}\right)} + V(\varphi,p) \, ,
\end{align}
where the first term can be interpreted as an effective kinetic energy $T(\varphi,p)$ and the second term as the potential the field evolves in.

With the Hamiltonian at hand, we can now explore the global stability of the field. If there exists a (constant) $p$-time $p_0$ such that, $ \forall p \geq p_0 $,
\begin{align}
\label{cnd}
& \left[\beta(1 - 3\varpi) + 2\widetilde{m}^2\right] \geq 0 \, , \nn \\
& \Leftrightarrow m^2 \geq \, \, |\beta| (1-3\varpi)\frac{\kappa \widetilde{\rho}_{0}}{2} e^{-3(1+\varpi)p} \, ,
\end{align}
then $V(\varphi,p)$ has a global minimum in $\varphi \, \, \forall p \geq p_0 $. This can easily be checked by plotting $V(\varphi,p)$ as a function of $p$ or $\varphi$. When this is the case, we say that the potential is \emph{stabilizing}.

For most of the history of the Universe, such a $p_{0}$ does exist and the potential is stabilizing. This is because $1 + \varpi>0$ (except during dark-energy domination and inflation). However, in very late-time cosmology (during dark-energy domination), $\varpi = -1$. In this case, Eq.~\eqref{cnd} with $\varpi=-1$ tells us that the potential is stabilizing if and only if
\begin{align}
\label{stab}
m^2 \geq \, \, 2 |\beta| \Lambda \, ,
\end{align}
where $\Lambda$ is the cosmological constant.

Let us now restrict attention to the dark energy-dominated era and use $\mathcal{H}$ to infer the behavior of the field as $p\rightarrow + \infty$. When the potential is stabilizing, 
\begin{enumerate}
\item $\mathcal{H}(\varphi,\varphi')$ \footnote{During dark energy-dominated era, the Hamiltonian does not depend explicitly on $p$ because $\widetilde{m}$ and $\rho$ are constants.}$ > 0 \, \, \forall \, \, \{\varphi,\varphi'\} \neq \{0,0\} \label{cond1} \, ,$
\item $\mathcal{H}(0, 0) = 0  \label{cond2} \, .$
\end{enumerate}
These conditions mean that the energy of the field is always bounded from below (by zero) and that it reaches its minimum at the phase space point $\{\varphi=0,\varphi'=0\}$. Moreover, Eq.~\eqref{eq:Hamiltonian} and Eq.~\eqref{eq_sta} also tell us that 
\begin{align}
\label{decay}
\frac{d \mathcal{H}}{dp} = - 2 \varphi'^2 \leq 0\,,
\end{align} 
and thus, $\mathcal{H}$ is semi-negative definite. 

In order to obtain Eq.~\eqref{decay}, one has to be careful. The potential used in Eq.~\eqref{eq:Hamiltonian} was obtained neglecting the $\varphi$ dependence of $\rho$, and thus, one should use the same approximation to obtain Eq.~\eqref{decay}. The above result, however, does not actually rest on this approximation. One can obtain Eq.~\eqref{decay} without knowing the exact analytic solution to Eq.~\eqref{EulerLagrange} for the potential. To do so, one simply uses the chain rule: $dH/dp = dV/dp + dT/dp = (dV/d\phi) \phi' + dT/dp$, and then inserts Eq.~\eqref{EulerLagrange} for the derivative of the potential with respect to the field. Doing so, one then arrives at Eq.~\eqref{decay} generically during the dark energy-dominated era. 

With this at hand, we can now determine the late-time stability features of the scalar field and explain why the amplitude of the derivative of the scalar field begins to decrease only during the dark energy-dominated era. Let us define $\Xi = \{(\varphi,\varphi'): \, \varphi'=0\}$ as the set of points at which the derivative of $\mathcal{H}$ vanishes. The only orbit of $\varphi$ contained in $\Xi$ is defined by the phase space point $\{\varphi(p),\varphi'(p)\}=\{0,0\}$. In fact, if $\{\varphi,\varphi'\} \in \Xi$ with $\varphi \neq 0$, then Eq.~\eqref{eq_sta} implies that $\varphi'' \neq 0$, and thus, the orbit leaves $\Xi$. Thanks to the first Lyapunov stability theorem and the Krasovskii-LaSalle invariance principle, we conclude that $ \{ \varphi=0,\varphi'=0 \}$ is \emph{asymptotically stable}. Therefore, at late times, the $\varphi$ field must flow to this orbit, settling down at the global minimum of the potential. In other words, due to the effective drag force, the energy of the scalar field decreases with time, forcing the field to settle down to the phase space point $ \{ \varphi=0,\varphi'=0 \}$ as $p \rightarrow + \infty$. This argument does not hold during radiation-dominated era or matter-dominated era because the inequality ${d \mathcal{H}}/{dp} \leq 0$ also does not hold. Furthermore, when the inequality in Eq.~\eqref{stab} is not satisfied, the field evolves in a potential that is not bounded from below, which forces it to diverge as $p \rightarrow + \infty$; this is precisely what happens in the massless case.

%-----------------
\subsubsection{WKB Approximation}

Let us now attempt to understand the approximate evolution of the the envelope of $\varphi(p)$ and the frequency of the oscillations using a well-known technique in multiple scale analysis: the WKB approximation. The original evolution equation for the scalar field in massive DEF theory [Eq.~\eqref{cosmo_KG}] expressed in the variable $p$ is
\begin{align}
\label{KG_p}
H^2 \varphi'' + \left(H H' + 3 H^2\right)\varphi' = -\left[\frac{1}{2}\beta(1-3\varpi)\kappa\rho + m^2\right]\varphi\,,
\end{align}
where recall the Hubble parameter satisfies the modified Friedmann equation [Eq.~\eqref{firstfrid}], which we rewrite below with the mass potential for convenience:
\begin{align}
\label{fried}
3 H^2\left(1-\frac{1}{3}\varphi'^2\right) = \kappa\rho + m^2 \varphi^2\,.
\end{align} 
From Eqs.~\eqref{KG_p} and~\eqref{fried}, the field is effectively massless if and only if $\kappa\rho \gg m^2 \varphi^2$~\citep{Damour_cosmo}. On the other hand, when the energy density has decayed enough so that $\kappa\rho \sim m^2 \varphi^2$, then the mass term is non-negligible.

Let us first consider the late-time behavior of the field, where $\varphi \ll 1$, and the regime $\kappa\rho \sim m^2 \varphi^2$. We then make the (WKB-inspired) substitutions 
\begin{align}
\varphi &\rightarrow \varepsilon \varphi\,, 
\quad 
\kappa\rho \rightarrow \varepsilon^2 \kappa\rho\,,
\quad
H \rightarrow \varepsilon H\,, 
\end{align}
with $\varepsilon \ll 1$, and Eq.~\eqref{KG_p} becomes
\begin{align}
\label{WKB}
H^2 \varphi'' + \left(H H' + 3 H^2\right)\varphi' = -\left[\frac{1}{2}\beta(1-3\varpi)\kappa\rho + \frac{m^2}{\varepsilon^2}\right]\varphi\,,
\end{align}
which is well-adapted to a WKB analysis due to the factor of $\varepsilon^{2}$ in the denominator of the last term. We therefore look for a WKB solution of the form 
\begin{align}
\varphi = \operatorname{Re}\left[{e^{\frac{1}{\delta}\sum \delta^n S_n}} \right]\, ,
\end{align}
where $\delta \ll 1$ and $S_{n}$ is in principle complex. A quick dominant balance calculation shows that we need $\delta = \varepsilon$. In other words, when $\varphi = \mathcal{O}(\varepsilon)$ with $\varepsilon \ll 1$, there is a short time scale of order $\mathcal{O}(1/\varepsilon)$ that naturally arises in the evolution of the field, as shown in the oscillatory behavior of the numerical solution. 

With this WKB ansatz, the evolution equation reduces to a set of ordinary differential equations (order by order in $\varepsilon)$: 
\begin{align}
\label{S0}
(S_0')^2 & = -\frac{m^2}{H^2}\,, \\
\label{S1}
2 S_1' & = -3 - \frac{1}{H}H'-\frac{S_0''}{S_0'}\,.
\end{align}
The last two terms of the right hand side of Eq.~\eqref{S1} cancel, upon inserting Eq.~\eqref{S0} in~\eqref{S1}. Solving these equations, the solution becomes
\begin{align}
\label{varphi_WKB}
\varphi(p) = e^K e^{-\frac{3}{2}p} \cos{\left[\int \frac{m}{\varepsilon H}dp\right]} + {\cal{O}}(\varepsilon)\,,
\end{align}
where $K$ is a constant\footnote{Here, $K \in\mathcal{R}$ and we ignore any initial phase in the oscillations of the field.}. 

In order to obtain a closed expression for $\varphi$ in terms purely of $p$-time, we now need to solve for $\varepsilon H$ at second order in $\varepsilon$. This can be achieved by inserting Eq.~\eqref{varphi_WKB} in~\eqref{fried} and expanding the Hubble parameter as $H = H_0 + \varepsilon H_1$. Doing so, we obtain
\begin{align}
\label{dust}
3 H_0^{2} & = \kappa \rho + m^2 e^{2 K} e^{-3p} \, , \\
\label{eq:H1}
H_1 & = \frac{3}{4}e^{-3p} e^{2K} m \sin{\left[2 \int \frac{m}{\varepsilon H}dp\right]} \,.
\end{align}
Equation~\eqref{eq:H1} shows that $H_1$ is a rapidly oscillatory function, and the derivative of its phase is always non-zero. Therefore, the integral of $H_{1}$ over $p$ is much less than unity, and we can safely replace $H$ by $H_0$ in the denominator of the integrand of $H_1$. One can check that this expansion is valid as long as
\begin{align}
\frac{\varepsilon H_1}{H_0} \ll 1 \quad \Rightarrow \quad  \varepsilon e^K e^{-\frac{3}{2}p} \ll 1 \, ,
\end{align}
i.e.~as long as the amplitude of the field is very small compared to one.

Combining all of these results, we obtain the late-time behavior of the scalar in the $\kappa\rho \sim m^2 \varphi^2$ regime: 
\begin{align}
\label{scalarWKB}
\varepsilon \varphi(p) = \operatorname{Re}\left[\overline{\varphi}_0 e^{-\frac{3}{2}p} e^{\pm i \int \frac{m}{\varepsilon H_0}- \frac{m H_1}{H_0^{2}}dp}\right] + {\cal{O}}(\varepsilon^2)\,,
\end{align}
where we have defined $e^K \varepsilon = \overline{\varphi}_0$, as the amplitude of the scalar field today. Similarly, we obtain the amplitude of the frequency of the scalar field\footnote{The limit used here makes sense rigorously only if there were only baryonic matter in the universe.}:
\begin{align}
\lim \limits_{p\rightarrow\infty} |\varphi'| \, = \sqrt{\frac{3}{1 + \kappa \widetilde{\rho}_{0}/(m^2 \overline{\varphi}_0^2)}} \,.
\end{align}
Clearly, the scalar field presents an exponentially-damped oscillatory behavior, while its frequency becomes approximately $\sqrt{3}$ if the cosmological energy density associated with the scalar field is much larger than that associated with usual matter, just as we found in the full numerical solution of Sec.~\ref{num}. 

The approximate solution found in the regime $\kappa\rho \sim m^2 \varphi^2$ is uniformly valid during the matter-dominated era, but not during the dark energy-dominated era or the radiation-dominated era. This is because during matter domination the potential energy of the field, which is proportional to $\varphi^2$, and the matter energy density both decay as $e^{-3p}$. But during the dark energy-dominated era, the energy density associated with $\Lambda$ is roughly constant (in the Einstein frame), and thus the inequality $(\kappa \rho)_{\Lambda} = \Lambda \gg m^2 \varphi^2$ eventually becomes satisfied. In fact, this inequality becomes stronger and stronger as the field evolves and continues to decay. 

Let us then consider the dark energy-dominated case with $\Lambda \gg m^2 \varphi^2$ and linearize the equation of motion [Eq.~\eqref{field_tot}] in the amplitude of the field:
\begin{align}
\varphi'' + 3 \varphi' + 3\left(2\beta + \frac{m^2}{\L}\right)\varphi = 0 \,.
\end{align}
The analysis of this equation leads to the same conclusions we arrived at in Sec.~\ref{subsec:H}, i.e.~one must have $m^2 \geq 2  |\beta| \L$ for the amplitude of the field to decay to zero as $p \rightarrow \infty$. In the limit where $m^2 \gg \L$ (which is compatible with the previous assumption $\Lambda \gg m^2 \varphi^2$ only for very small values of $\varphi$) and for $\beta$ of order unity, we can integrate the previous equation to find 
\begin{align}
\label{field_lambda}
\varphi(p) = \overline{\varphi}_0 e^{-\frac{3}{2}p}\cos\left({\frac{3 m^2}{\L}p}\right)\,.
\end{align}
Again, we see that this is also the same approximate behavior we found in the exact numerical solution presented in Sec.~\ref{num}. 

Equation ~\eqref{field_lambda} reveal that the amplitude of the derivative of the scalar field decays during dark energy-domination (at least when the potential energy of the field is small compared to the cosmological constant). This is consistent with our numerical results and confirms that, in the analytical treatment of this subsection, we were justified in neglecting terms that scale as $\varphi'^2$ in the field equation. Moreover, there is a continuous mapping between the two solution [Eqs.~\eqref{field_lambda} and~\eqref{scalarWKB}], suggesting that the matching of their asymptotic behaviors would provide a continuous solution at the matter-domination/dark energy-domination interface. 

As we found in Sec.~\ref{num}, during the radiation-dominated era the field starts decaying when its potential energy and the radiation energy density are of the same order. In this regime, the approximate solution of Eq.~\eqref{field_lambda} is also in very good agreement with the numerical simulations. When the potential energy of the field begins to dominate over the radiation energy density, our approximation breaks down and the characteristic time of variation of $\varphi$ decreases slightly. Our analysis does not capture this feature, but fortunately this behavior is not relevant when considering Solar System tests at late times.  

Before concluding this subsection, let us point out that the term proportional to $\beta$ in the field equation [Eq.~\eqref{firstfrid}] is negligible. This is not a surprise because when $\varphi \ll 1$ at late time, then the Einstein and the Jordan frames become approximately equal. Since our simulations show that the amplitude of the field does decay exponentially, this approximation is always true after a certain amount of cosmological time (that depends on the mass the field) has elapsed. Thus, the parameter $\beta$ weakly influences the behavior of the field in late-time cosmology as long as $m^2 \gg \Lambda$.

%------------------------------------------------------------------------------------
\subsection{Late-Time Cosmology}

One may be concerned that the oscillations presented in the previous subsection are inconsistent with late-time cosmology, i.e., with observations at small redshift. In this section, we show that massive DEF theory in fact embeds consistently in this context. As we will see below, the highly-oscillatory scalar field behaves like cold-dark matter, and thus, its impact on the evolution of the universe is consistent with cosmological observations. 

\begin{figure}[tb]
\begin{center}
\includegraphics[width=\columnwidth,clip=true]{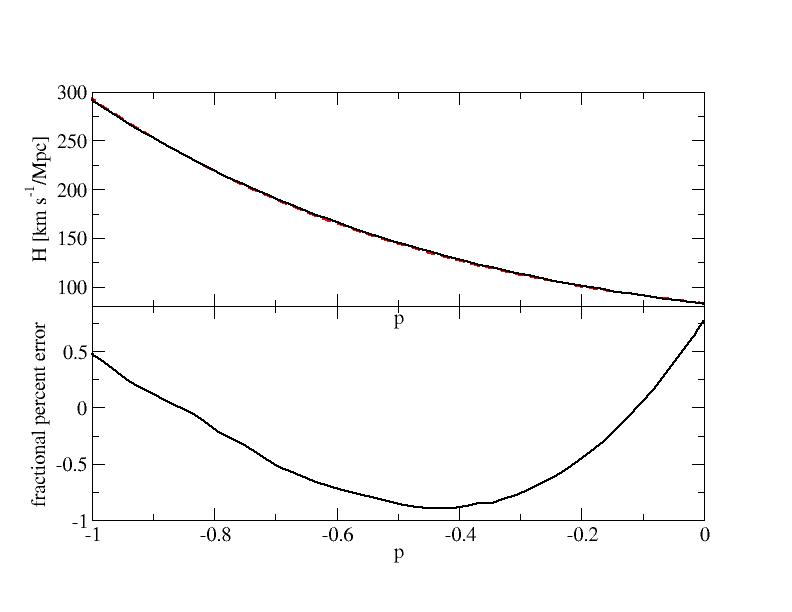}
\caption{\label{fig:hubble}~(Color online) Hubble parameter as a function of $p$-time for $\beta=-4.5$, $m=5.8 \times10^{-28} \text{eV}$ and $\varphi_0=1$. Top: The Hubble parameter is computed first by solving the field equations numerically (dashed red line), and then fitting this data to the analytical approximate model of Eq.~\eqref{hubble} (black line). Bottom: Fractional percent difference between the numerical and the analytical result. Observe that the error between the numerical solution and the analytic expression is at most approximately $1 \%$.}
\end{center}
\end{figure}

Given the generic result of the numerical simulations presented in the previous subsection, the scalar field is well damped at late times. This implies that there is effectively no difference between the Einstein and the Jordan frames. Of course, for this to be true the massive transition, i.e., the transition between the quiescent phase and the oscillatory phase, must occur before dark energy domination, which translates roughly to requiring $ m^2 \gg \Lambda $. In this regime, the first Friedmann equation becomes
\begin{align}
\label{hubble}
3 H^2 = &  \frac{\left(\Lambda + m^2 \varphi^2 \right)}{\left(1-\frac{\varphi'^2}{3}\right)} 
     \simeq  \left(\Lambda + m^2 \overline{\varphi}_0^2 e^{-3p}\right) \,,
\end{align}
the last expression being valid at leading order in the amplitude of the scalar field when $\Lambda \sim m^2 \overline{\varphi}_0^2 e^{-3p}$ or when $\Lambda \gg m^2 \overline{\varphi}_0^2 e^{-3p}$. The next term in the perturbative expansion of $H$ in power of $\overline{\varphi}_0$ is given in Eq.~\eqref{dust}. The scalar field therefore behaves as dust on cosmological scales (as shown by the exponential decay with a charateristic time of $1/3$ of the contribution of the scalar field to the scale factor) and its oscillatory behaviour only appears in the scale factor perturbatively. This is confirmed by our numerical simulations as shown in Fig.~\ref{fig:hubble}, where the dashed red curve is the numerical evolution of the Hubble parameter, and the black continuous curve is a fit to this data with the model of Eq.~\eqref{hubble}.

\begin{figure}[tb]
\begin{center}
\includegraphics[width=\columnwidth,clip=true]{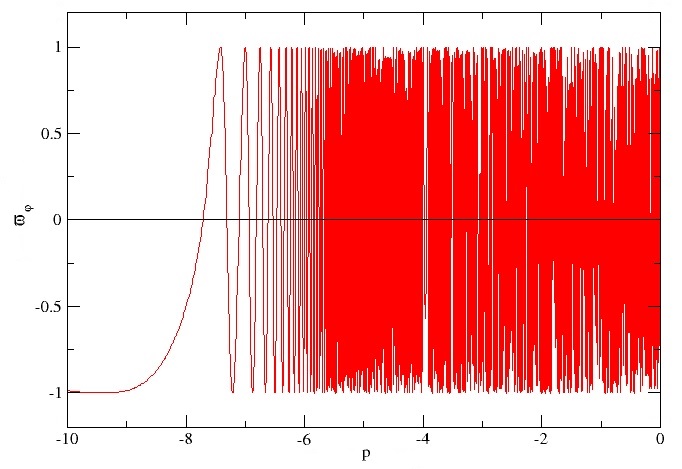}
\caption{\label{fig:EOS} Evolution of the equation of state of the scalar field as a function of $p$-time, obtained by numerically solving the field equation with $\beta=-4.5$, $m=5.8 \times 10^{-28} \text{eV}$ and $\varphi_0=1$.Observe the equation of state averages to zero.}
\end{center}
\end{figure}

This result should not be a surprise because the Einstein frame equation of state of the field is given by
\begin{align}
\varpi_{\varphi} & = \frac{P_\varphi}{\rho_\varphi} = \frac{\dot{\varphi}^2 - m^2 \varphi^2}{\dot{\varphi}^2 + m^2 \varphi^2}\,,
\end{align}
which averages to 0, as we can see on Fig.~\ref{fig:EOS}. In the previous equation, $P_\varphi$ and $\rho_\varphi$ are the effective pressure and energy density associated with the scalar field respectively. These quantities can be read off from the modified Friedmann equation [Eq.~\eqref{hubble}]. 

Such a scalar field thus could in principle act like cold dark matter on cosmological scales. In that sense, our results are closely related to those of the scalar field dark matter model~\cite{SFDM}. This model is obtained by setting $\beta = 0$ in the massive DEF model, as e.g.~explained in detail in~\cite{SFDM}. Here, we do not try to constrain any parameter of the theory by matching the coefficient $\overline{\varphi}_0$ to the dark matter $\Lambda CDM$ parameter, as it seems too restrictive to postulate the absence of any other type of dark matter particles/fields, such as weakly interacting massive particles. The main point here is that the late-time behavior of the scalar field, and its impact on the evolution of the scale factor, is consistent with cosmological observations. 

%%%%%%%%%%%%%%%%%%%%%%%%%%%%%%%%%%%
\section{Solar System Constraints}
\label{sec:sol}

In this section, we carry out a weak-field analysis of massive DEF theory with the aim to calculate Solar System observables that we can then compare against observations. As we shall find, the weak-field analysis is complicated by the persisting time-dependence of the scalar field in the Solar System, which forces us to carry out a time-dependent perturbative expansion.

%--------------------------------------------------
\subsection{Scalar Field Behavior in the Solar System}

In order to understand what happens on Solar System scales, let us first derive the solution to the modified Klein-Gordon equation around a given body, neglecting the curvature of spacetime. This equation becomes
\begin{align}
\label{KG_simple}
\square_{\eta} \varphi = m^2 \varphi + 4 \pi G \beta \; \varphi \; e^{2\beta\varphi^2} \widetilde{\rho}\,,
\end{align}
where $\square_{\eta}$ is the D'alembertian operator in flat spacetime, and where we have assumed that $\widetilde{\rho} \gg \widetilde{P}$, with $\widetilde{\rho}$ and $\widetilde{P}$ the energy density and the pressure of the body creating the perturbation respectively. Let us approximate the matter distribution as spherically symmetric, and so let us use spatial spherical coordinates $\{ r, \theta, \phi \}$.

An important feature of the results presented in the previous sections is that the cosmological scalar field oscillates with conformal time $p$ as given by Eq.~\eqref{scalarWKB}, with (angular) frequency $\omega_p$ today of
\begin{align}
\omega_p = \frac{m}{H_0}\,,
\end{align}
where $H_0$ is the Einstein frame Hubble constant. Consequently, the cosmological field oscillates with the coordinate time $t$ of the FLRW metric (which is the proper time of observers at rest in this spacetime) with an angular frequency of
\begin{align}
\label{puls}
\omega = m
\end{align}
This is, of course, only true today, i.e.~in the limit $\overline{\varphi}_0 \ll 1$, where $\overline{\varphi}_0$ is the amplitude of the cosmological scalar field. Recall that this condition is a consequence of the cosmological evolution of the scalar field in the $m^2\gg \L \sim H_0^2$ regime, which we focus on here in order to remain consistent with the cosmological constraint obtained in Eq.~\eqref{stab} for $\beta$ of order unity. 

This behavior imposes a (cosmological) boundary condition for the scalar field in the Solar System, namely 
\begin{align}
\label{eq:cosm-BC}
\lim_{r\to+\infty} \varphi(r,t) = \overline{\varphi}_0 \cos{(m\,t)}\,.
\end{align}
One cannot therefore assume that the scalar field asymptotes to a constant at spatial infinity, like one does in massless DEF theory. In fact, for the values of the mass usually of interest to massive scalars in cosmology and in the scalar field dark matter model (e.g. see \cite{SFDM}), i.e., when $m \sim 10^{-20} {\rm{eV}}$, the period of oscillation is roughly 
\begin{align}
\label{ODG}
\frac{\varphi}{\dot{\varphi}} \approx 0.01 \; {\rm{yrs}} \;  \left(\frac{10^{-20} {\rm{eV}}}{m}\right)\,. 
\end{align}
Solar System experiment with observation times larger than a few days would then be sensitive to the temporal variations of the scalar field. Even though cosmological observable quantities, such as the scale factor, evolve on time scales of order $H_0^{-1}$, the cosmological scalar field varies on a much shorter time scale that needs to be taken into account when dealing with its influence on astrophysical observables. 

Given the above considerations, let us search for an approximate solution to our toy problem in Eq.~\eqref{KG_simple} of the form
\begin{align}
\varphi(r,t) = f(r)\cos{(m\,t)}\,,
\end{align}
with the boundary condition $f(r) \to \overline{\varphi}_0$ as $r \to +\infty$.  Inserting this ansatz in Eq.~\eqref{KG_simple} leads to
\begin{align}
\frac{1}{r^2}\partial_r\left(r^2\partial_r f\right) = 4 \pi G \beta f \widetilde{\rho}\,,
\end{align}
which is simply the Poisson equation, i.e., a \emph{massless} equation. The solution to this equation, of course, depends on the density profile, but we expect, outside of the body, a solution of the form $f(r) = \overline{\varphi}_0 + \Pi/r$, where $\Pi \propto M$ is an integration constant and where $M$ is the mass of the astrophysical body, leading to the full solution
\begin{align}
\varphi(r,t) = \left( \overline{\varphi}_0 + \frac{\Pi}{r} \right) \cos{(m\,t)}\,\,.
\end{align}
In the next section, we will use post-Newtonian theory to find the proper solution for the scalar field, assuming a weak-field, constant-density astrophysical body, and we will find that indeed it takes the form presented above.

This solution appears to be in complete conflict with our standard expectations of massive DEF theory in the Solar System. When solving the Klein-Gordon equation for a massive scalar field in a \emph{stationary}, i.e.~time-independent, and spherically symmetric geometry that is asymptotically flat, one finds a Yukawa-type potential
\begin{align}
\label{eq:bad-sol}
\varphi(r) \propto \frac{M}{r}  e^{-m r}\,.
\end{align} 
for large radii. This solution comes about because we required \emph{a priori} that $\varphi$ be time-independent. This requirement forces any time derivatives in the left-hand side of Eq.~\eqref{KG_simple} to vanish, which then means that $\varphi(r)$ must by itself reproduce the right-hand side of Eq.~\eqref{KG_simple}. The latter is not necessary when $\varphi$ is allowed to also be time-dependent, since then the time derivatives in the left-hand side of Eq.~\eqref{KG_simple} can cancel the term proportional to $m^{2}$ on the right-hand side of this equation.

Requiring that the scalar field be time-independent creates a problem when requiring that the scalar field matches smoothly between its cosmological evolution and its Solar System behavior. If one forces the field to be static, then the solution to its evolution equation [Eq.~\eqref{eq:bad-sol}] is Yukawa (exponentially) suppressed when $r > 1/m$, which in particular implies that $\varphi \to 0$ as $r \to + \infty$. Of course, we immediately recognize that this boundary condition is in conflict with the boundary condition imposed by the cosmological evolution of the scalar field [Eq.~\eqref{eq:cosm-BC}]. This problem disappears when allowing for the scalar field to be time-dependent, since the cosmological boundary condition can then be easily satisfied. 

The need to account for the cosmological time-dependence of scalar fields goes beyond massive DEF theory, and in fact, it is applicable to a much wider class of models. For example, we showed in Sec.~\ref{sec:cosmo} that as long as $\beta$ is not very large compared to unity, the behavior of the cosmological scalar field at late times does not depend on $\beta$ . Consequently, the same type of scalar field behavior could be expected when adding simply a massive scalar to the stress-energy tensor (without the Brans-Dicke non-minimal coupling) and studying its excitations on astrophysical scales. 
%NY: Not sure what you meant by this: ``just requires the field to be homogeneous on sufficiently large scales - no matter how small this background value could be. In fact, as we have shown ''
 
%------------------------------------------------------------------------------------
\subsection{Perturbative Decomposition}

With the previous toy problem under control, we now compute observable effects in the Solar System through a post-Newtonian analysis. We perform all calculations in the Einstein frame, rewriting only the end results in the Jordan frame to find observables. In the Einstein frame, we expand all quantities to first order around a fixed background:
\begin{align}
& \varphi = \varphi_{(0)} + \epsilon~ \varphi_{(1)}\,, \\
& T_{\mu\nu} = T^{(0)}_{\mu\nu} + \epsilon~ T^{(1)}_{\mu\nu}\,, \\
& g_{\mu\nu} = g^{(0)}_{\mu\nu} + \epsilon~ g^{(1)}_{\mu\nu}\,, 
\end{align}
where the subscripts (or superscripts) $(0)$ and $(1)$ refer to background or first-order quantities respectively. We use $\epsilon \ll 1$ as an order counting parameter for the metric perturbation -- namely, in the Solar System $\epsilon = {\cal{O}}(G M / R_{\SSs})$ with $GM$ the Schwarzschild radius of the Sun and $R_{\SSs}$ the typical size of the Solar System. We restrict ourselves to the study of a spherically symmetric background and perturbations, but the results found here can be easily extended to more generic scenarios. Moreover, when carrying out order-of-magnitude calculations, we treat $\beta$ as a quantity of order unity for simplicity, although our equations do not rely on this assumption. 

The background scalar field $\varphi_{(0)}$ must behave as prescribed by its cosmological evolution, and thus, $\varphi_{(0)}$ is at leading order spatially homogeneous, isotropic and purely time dependent\footnote{We shall see later on that multiple scale analysis has to be used in order to solve the field equations consistently, leading to a first order spatial dependence of $\varphi_{(0)}$.}. Consequently, any spatial derivatives of the scalar field are treated as first order quantities in the sense that $\varphi^{(0)}$ varies only on length scales proportional to $R_{\SSs}/\epsilon$. The first order part of the scalar field is expected to be sourced by a weak-field, astrophysical source, and thus, it is allowed to depend both on time and spatial coordinates. 

The perturbed stress-energy tensor represents a weak-field, spherically-symmetric astrophysical matter source -- namely, a source for which the ratio between its internal pressure and energy density is of $P/\rho = \mathcal{O}(\epsilon)$, and for which the Virial theorem guarantees the characteristic velocities to be $v =  \mathcal{O}(\sqrt{\epsilon})$. Consequently, we expect the first order metric perturbation to be sourced only by the energy density of the body creating the perturbation. Moreover, on Solar System scales, we assume that there is no background $T^{(0)}_{\mu \nu}$. On scales much larger than the Solar System, the background stress energy tensor is the cosmological one $T_{\mu \nu}^{(0)} = T_{\mu\nu}^{\rm cosmo}$, where this includes the cosmological constant, the CMB and every cosmological sources of stress-energy, apart from the scalar field.

\begin{figure}[htb]
\begin{center}
\includegraphics[width=\columnwidth,clip=true]{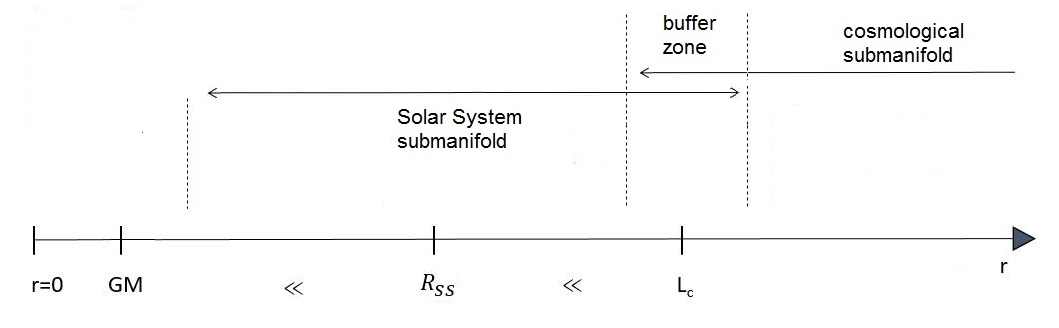}
\caption{\label{fig:scales} Schematic representation of a spacelike hypersurface $t = \text{cst}$ of the spacetime, the two sub-manifolds we consider, and the different scales at play in our perturbative scheme. The region $r \gg L_C$ is dominated by cosmology and the behavior of the field is given in Sec.~\ref{sec:cosmo}. We are interested in computing observables in the Solar System submanifold, which is characterized by length scales of ${\cal{O}}(R_{\SSs})$. 
}
\end{center}
\end{figure}

We decompose the spacetime into two sub-manifolds of co-dimension zero: a cosmological one, located at supra-solar distances, and a Solar System one, located at smaller distances (but at distances still large relative to the Schwarzschild radius of the astrophysical body creating the perturbation). Section~\ref{sec:cosmo} provides the behavior of the scalar field and of the metric in the cosmological submanifold, where all fields are spatially homogeneous. We are thus here interested in finding the behavior of all fields in the Solar System submanifold, which must asymptotically match the cosmological behavior in some buffer zone characterized by a certain length scale $L_C$. We know that the characteristic length scale associated with the Newtonian potential is the Schwarzschild radius $G M$, whereas the length scale associated with cosmology is $H_0^{-1}$. In the following, we define the buffer zone through a geometric average between these two scales, i.e.~ $L_C \sim (G M)^{\eta}H_0^{-1 + \eta}$ with $\eta < 1$. A convenient choice is to pick $\eta = 1/2$ for which we have $L_C \gg R_{\SSs}$ -- namely the dynamics in the Solar System are, as one expects, dictated by sub-cosmological length scales. An schematic representations of the different scales at play is given in Fig.~\ref{fig:scales}.

Since the background spacetime is sourced by the background scalar field $\varphi_{(0)}$, which we already argued must be spatially homogeneous and isotropic, using insight from our cosmological analysis, we search for solutions to the modified Einstein field equations that are conformally related to a perturbation of the Minkowski metric. We then consider the ansatz 
\begin{align}
g_{\mu\nu} & = e^{2 \Omega} \overline{g}_{\mu\nu}= e^{2 \Omega}\left(\eta_{\mu\nu} + \epsilon \; h_{\mu\nu}\right)\,,
\end{align}
where clearly then 
\begin{align}
g_{\mu \nu}^{(0)} &= e^{2 \Omega} \; \eta_{\mu \nu}\,,
\\
g_{\mu \nu}^{(1)} &= e^{2 \Omega} \; h_{\mu \nu}\,,
\end{align}
where $\Omega$ depends on time only. We further work in harmonic coordinates associated to $\overline{g}_{\mu \nu}$, defined by the gauge condition
\begin{align}
\overline{g}^{\mu\nu} \overline{\Gamma}^\alpha_{\mu\nu} = 0\,,
\end{align}
with $\overline{\Gamma}$ the Christoffel symbol associated with $\overline{g}_{\mu \nu}$. Equivalently, at linear level, this condition can be rewritten as
\begin{align}
\partial_\mu h^\mu_\lambda - \frac{1}{2}\partial_\lambda h = 0\,.
\end{align}
where the indices are raised and lowered by the Minkowski metric and where $ h = \eta^{\mu\nu} h_{\mu\nu}$ is the trace of the metric perturbation. In this (harmonic) gauge,  $x^i$ with $i \in (1,3)$ denotes spatial coordinates and $\tau$ denotes the temporal coordinate.

Before proceeding, let us stress that the following computation rests on some \emph{strong} assumptions. The most important one is that both the energy density and the deviations to Minkowski are treated as perturbations. This prevents the emergence of any type of nonlinear coupling between the perturbed field and any of these quantities. In the strong field regime, where these assumptions do not hold, the ansatz we use may not be the correct one and the solution of the field equations may be more involved than the one we present in the following analysis.

%------------------------------------------------------------------------------------
\subsection{Zeroth Order Evolution}

At zeroth order, and neglecting spatial variations of the conformal factor $\Omega$, the field equations reduce to the modified Friedmann equations and the Klein-Gordon equation. This then immediately implies that 
\begin{align}
e^{2\Omega} = a^2(\tau) \, ,
\end{align}
where $a(\tau)$ is the usual Einstein frame cosmological scale factor. Moreover, this also implies that the background spacetime is the FLRW metric
\begin{align}
ds^2_{(0)} = -dt^2 + a(t)^2 \left( dr^2 + r^2 d\Omega^2 \right) \, ,
\end{align}
in spherically symmetric coordinates with $dt = a(\tau)d\tau$.

We require that our theory is cosmologically consistent, i.e.~we assume that the predicted Hubble constant is indeed the observed one $H_0$. Then, in the limit where $m \gg H_0 \sim 10^{-33}$ eV, the Klein-Gordon equation becomes
\begin{align}
\left(\partial_{\tau}^2 \varphi_{(0)} + 2 {\cal{H}} \partial_\tau \varphi_{(0)} \right) + m^2 \varphi_{(0)} =  \beta \varphi_{(0)} T^{\rm cosmo}\,,
\end{align}
where ${\cal{H}} = \partial_{\tau} a/a$ is the conformal Hubble parameter and $T^{\rm cosmo} = \eta^{\alpha\beta}T^{\rm cosmo}_{\alpha\beta}$ is the trace of the cosmological stress-energy tensor, simplifies to
\begin{align}
\partial_\tau^2\varphi_{(0)} + m^2 \varphi_{(0)} = 0\,,
\end{align}
since the trace of the stress-energy tensor if of ${\cal{O}}(H_0)$. This simplification implies we treat all time-dependent contributions to $\varphi_{(0)}$ that evolve on time scales comparable to $H_0^{-1}$ as constants. The background scalar field is then
\begin{align}
\label{back}
\varphi_{(0)}(\vec{r},t) = A(\vec{R})e^{i m \tau} + A^*(\vec{R})e^{-i m \tau} \, ,
\end{align}
where the length scale $R$ is assumed large relative to the size of the system under consideration, i.e.~relative to the size of the Solar System $R_{\SSs}$.

For later convenience, let us define a new set of coordinates $(t,u,\Omega)$, where $u = a(t) r$ is a co-moving radial coordinate. In this coordinates, the FLRW line element becomes
\begin{align}
ds^2_{0} = -\left\{1 - [H(t)u]^2\right\}dt^2 - 2H(t)u dt du + du^2 + u^2 d\Omega^2 \, ,
\end{align}
where $H(t) = \partial_{t} a/a$ is the Hubble parameter. Henceforth, we assume that $H(t)R_{\SSs} \ll \varepsilon$, which allows us to solve for first-order quantities, taking the background to be Minkowski, which is a legitimate approximation on Solar System scales. 

%------------------------------------------------------------------------------------
\subsection{First-Order Evolution}
Henceforth, we work with coordinates $\{t,x^i\}$ and we neglect the influence of the scale factor on Solar System scales. We begin by decomposing the stress-energy tensor, then finding the evolution equations and finally solving them to first-order. 

%-----------------
\subsubsection{Perturbed Stress-Energy Tensor}
We describe the astrophysical body creating the perturbation as a pressureless perfect fluid. Its stress-energy tensor is defined in the Jordan frame at first order by
\begin{align}
\widetilde{T}_{\mu\nu}^{(1)} = \epsilon \; \widetilde{\rho} \; u_\mu u_\nu \,,
\end{align}
with the normalization condition $\widetilde{g}_{\mu\nu}^{(0)}u^\mu u^\nu = -1$ where $\widetilde{g}_{\mu\nu}$ refers to the Jordan frame metric and $\widetilde{\rho}$ is the Jordan frame $\epsilon$-normalized energy density.

From Eq.~\eqref{perfect_fluid}, in the Einstein frame, the perturbation is also described by a perfect fluid with an $\epsilon$-normalized density $\rho$ defined by
\begin{align}
\rho = e^{2\beta\varphi_{(0)}^2}\widetilde{\rho}\,.
\end{align}
Therefore, the corresponding stress energy in the Einstein frame takes the  form
\begin{align}
T_{\mu\nu}^{(1)} & = \epsilon \; e^{2 \beta  \varphi_{(0)}^2}\widetilde{\rho} \; \delta^0_\mu \delta^0_\nu\,,
\end{align}
and its trace is simply
\begin{align}
 T^{(1)} & = - \epsilon \; e^{2 \beta  \varphi_{(0)}^2} \; \widetilde{\rho}\,.
\end{align}

%-----------------
\subsubsection{Modified Field Equations and Klein Gordon Equation}

At first order and using harmonic coordinates, the modified field equations are
\begin{align}
- \epsilon \frac{1}{2}\eta^{\alpha\beta} \partial_\alpha \partial_\beta h_{\mu\nu} & = 8 \pi G \epsilon\left(T_{\mu\nu}^{(1)} - \frac{1}{2}T^{(1)} \eta_{\mu\nu}\right) \nn \\ & + 2 \partial_\mu (\varphi^{(0)} + \epsilon \varphi^{(1)})  \partial_\nu (\varphi^{(0)} + \epsilon \varphi^{(1)}) \nn \\ & - 2 \left(\partial_\t \varphi^{(0)}\right)^2 \delta^\t_\mu \delta^\t_\nu  + 4 m^2 \epsilon \varphi^{(0)} \varphi^{(1)} \eta_{\mu\nu} \, ,
\end{align}
These equations can be simplified greatly through the following reasoning. From the cosmological evolution of the scalar field in Sec.~\ref{sec:cosmo}, we know that $m \varphi_{(0)} \thicksim \partial_0 \varphi_{(0)} \thicksim H_0$, with $H_0$ the Hubble constant. Therefore, neglecting\footnote{This assumption is redundant with the one made in the previous section, which allowed us to use a Minkowski background.}  terms of ${\cal{O}}(R_S H_0)^2$ and ${\cal{O}}(R_S H_0)^2\varphi^{(1)}/\varphi^{(0)}$ leads to the following simplified set of equations 
\begin{align}
-\frac{1}{2}\nabla^2h_{00} &\simeq 4 \pi G e^{2 \beta \varphi_{(0)}^2} \widetilde{\rho}\,,  \\
-\frac{1}{2}\nabla^2h_{ii} &\simeq 4 \pi G e^{2 \beta \varphi_{(0)}^2} \widetilde{\rho}\,,  \\
\label{offdiag}
-\frac{1}{2}\nabla^2h_{ij} &\simeq 0\,, \text{ for } i\neq j\,, \\
-\epsilon \frac{1}{2}\nabla^2h_{0i} &= 2 \partial_\t \varphi^{(0)} \left( \partial_i \varphi^{(0)} + \epsilon \partial_i \varphi^{(1)} \right)\,,
\end{align}
where the right-hand side of Eq.~\eqref{offdiag} is technically non-zero, but would give rise to a solution that varies on time and length scales of order the Hubble radius, and so it can be neglected here. 

The structure of the modified field equations suggests that in the Einstein frame the perturbed line element takes the form
\begin{align}
ds^2 & = -\left[1 - \epsilon \chi(\vec{r},t)\right)]dt^2 + 2 \epsilon \Gamma_i dtdx^i 
\nn \\
&+ \left[1 + \epsilon \chi(\vec{r},t)\right]\delta_{ij}dx^idx^j \, ,
\end{align}
with 
\begin{align}
\label{chi}
\nabla^2 \chi(\vec{r},t) \simeq - 8 \pi G e^{2 \beta \varphi_{(0)}^2}\widetilde{\rho}(\vec{r}) \, , \\
\label{gamma}
\epsilon \nabla^2 \Gamma_i \simeq - 4 \partial_\tau \varphi_{(0)} \left(\partial_i \varphi_{(0)} + \epsilon \partial_i \varphi_{(1)} \right) \,.
\end{align}
Unlike in GR, first-order off-diagonal metric terms can potentially arise in massive DEF theory due to the non-vanishing of the time derivative of the cosmological field. If the latter varies with time only on cosmological scales (like the scale factor), as is the case in the massless DEF theory, then this term is automatically suppressed. As we have shown in Sec.~\ref{sec:cosmo}, however, this is not the case because the time derivative of the scalar field varies on scales of order the mass of the field, which can be much larger than the Hubble parameter today.  

Similarly, keeping only zeroth- and first-order terms and working in harmonic coordinates, the scalar field evolves according to  
\begin{align}
&\epsilon\eta^{\mu\nu}\partial_\mu\partial_\nu \varphi_{(1)} + \left(\eta^{\mu\nu} - \epsilon h^{\mu\nu}\right)\partial_\mu\partial_\nu \varphi_{(0)} \nn \\
&= \, m^2(\varphi_{(0)} + \epsilon \varphi_{(1)})  + 4 \epsilon \pi G \beta \varphi_{(0)} e^{2 \beta \varphi_{(0)}^2} \widetilde{\rho} \,.
\end{align}
Under the assumption that spatial derivative of $\varphi_{(0)}$ are first-order quantities, and using the zeroth-order evolution equation, we obtain
\begin{align}
-\epsilon \partial_t^2 \varphi_{(1)} + \epsilon \partial^i \partial_i \varphi_{(1)} + \partial^i \partial_i \varphi_{(0)} + \epsilon \chi m^2 \varphi_{(0)} \nn \\ = m^2 \epsilon \varphi_{(1)} + 4 \pi G \beta \epsilon\varphi_{(0)} e^{2 \beta \varphi_{(0)}^2} \widetilde{\rho}\,.
\end{align}
We render this equation dimensionless through (dimensionless) spatial coordinates $\sigma^i = x^i/R_{\SSs}$ and time coordinate $t' = m \, t$ to find
\begin{align}
\label{KG_dim}
-\epsilon (m R_{\SSs})^2 \partial_{t'}^2 \varphi_{(1)} + \epsilon \partial^i \partial_i \varphi_{(1)} + \partial^i \partial_i \varphi_{(0)} + \epsilon \chi R_{\SSs}^2 m^2 \varphi_{(0)} \nn \\ = \epsilon R_{\SSs}^2 m^2 \varphi_{(1)} + 4 \pi G \beta \epsilon R_{\SSs}^2 \varphi_{(0)} e^{2 \beta \varphi_{(0)}^2} \widetilde{\rho} \,.
\end{align}

%-----------------
\subsubsection{Metric and Scalar Field Solutions}

Now that we have found the first-order evolution equation for the scalar field [Eq.~\eqref{KG_dim}], let us solve it in the exterior of the matter source, which we assume is contained in a sphere of radius $R_{\source}$. Once more, we use that $\varphi_{(0)} \ll 1$ today and neglect terms that are proportional to $\varphi_{(0)}^3$ in the first-order equations. To solve Eq.~\eqref{KG_dim} with these assumptions, we first need to compute the solution to Eq.~\eqref{chi} neglecting the $\varphi_{(0)}^2$ contribution, and we obtain 
\begin{align}
\epsilon \chi(\vec{r},t') & =  2 G \left( 1 + \mathcal{O}(\varphi_{(0)}^2)\right) \int d^3\vec{r}_1 \frac{\widetilde{\rho}(\vec{r}_1)}{|\vec{r} - \vec{r}_1|} \,, \\
& = \frac{2 G M}{r} \left( 1 + \mathcal{O}(\varphi_{(0)}^2)\right) \,,
\label{eq:chi-approx}
\end{align}
with $M$ the enclosed mass, namely
\begin{align}
M = 4 \pi \int dr \, r^2 \epsilon \widetilde{\rho}\,.
\end{align}
Thus, for $\sigma \geq R_{source}/R_S$, Eq.~\eqref{KG_dim} becomes
\begin{align}
\label{KGBis}
-(m R_S)^2\partial_{t'}^2 \varphi_{(1)} + \frac{1}{\sigma^2}\partial_\sigma\left[\sigma^2\partial_\sigma \varphi_{(1)}\right] + \frac{1}{\epsilon\sigma^2}\partial_\sigma\left[\sigma^2\partial_\sigma \varphi_{(0)}\right] \nn \\ + \frac{2}{\sigma}(R_S m)^2 \varphi_{(0)} = (R_S m)^2 \varphi_{(1)} + 4 \pi G \beta R_S^2\varphi_{(0)} \widetilde{\rho}\,.
\end{align}

In the previous equation, the forcing term proportional to $\varphi_{(0)}/\sigma$ is resonant with the homogeneous solution for $\varphi_{(1)}$ and it is non-localized, possible leading to secular divergences of the field $\varphi_1$. This behavior is unphysical and it can be dealt with using multiple scales analysis as follows. We require that the function $A(\vec{R})$ appearing in Eq.~\eqref{back} be a function of the slowly-varying variable $\Sigma = \varepsilon \sigma$ only. Using this in Eq.~\eqref{KGBis}, we then find that 
\begin{align}
\label{multi}
\partial_\Sigma A + (R_S m)^2 A = 0\,,
\end{align}
whose solution is simply
\begin{align}
A(\sigma) = A e^{- \varepsilon (R_S m)^2 \sigma} \, ,
\end{align}
with $A \in \mathbb{R}$. This solution then cancels the resonant term that arises from the spatial derivative of the zeroth-order scalar field. The zeroth-order scalar field then becomes
\begin{align}
\varphi_{(0)}(\vec{r},t) = A e^{- m^2 GM r} \cos{(m \, t)}\,.
\end{align}

With this at hand, the first-order field equation reduces to
\begin{align}
-\partial_t^2 \varphi_{(1)} + \frac{1}{r^2}\partial_r\left[r^2\partial_r \varphi_{(1)}\right] =  m^2 \varphi_{(1)} + 4 \pi G \beta A(R) \widetilde{\rho}\,.
\label{eq:simped-DE}
\end{align}
We can find a solution to this equation through Fourier transform techniques. In the Fourier domain, Eq.~\eqref{eq:simped-DE} becomes
\begin{align}
\epsilon \hat{\varphi}_{(1)}(\vec{k},\omega) &= 4 \pi G \beta A(R)\sqrt{\frac{\pi}{2}}\frac{\epsilon\hat{\widetilde{\rho}}(\vec{k})}{-\vec{k}^2+\omega^2-m^2} 
\nn \\
&\times \left[\delta(\omega-m) + \delta(\omega+m)\right]\,,
\end{align}
where we have used the convention
\begin{align}
\hat{\varphi}_{(1)}(\vec{k},\omega) = \frac{1}{\sqrt{2 \pi}^4}\int d^3\vec{r} \,  dt \, e^{i(\omega t - \vec{k}\cdot \vec{r})} \varphi_{(1)}(\vec{r},t)\,.
\end{align}
Consequently,
\begin{align}
\varepsilon \varphi_{(1)}(\vec{r},t) & = \frac{1}{4 \varepsilon \pi^2}\int d^3 \vec{k} \,  d\omega \, e^{-i(\omega t - \vec{k} \cdot \vec{r})} \hat{\varphi}_{(1)}(\vec{k},\omega) \, , \nn \\
                     & = -\sqrt{2}\frac{G \beta A(R)}{\varepsilon \sqrt{\pi}}\cos{(m \, t )}\left[\int d^3\vec{k} \, \frac{e^{i \vec{k}\cdot\vec{r}}}{\vec{k}^2}\hat{\widetilde{\rho}}(\vec{k})\right] \, , \nn \\
%                     & = - \frac{G \beta A(R)}{\varepsilon 2 \pi^2}\cos{(m \, t )}\left[\int  d^3\vec{k} \, d^3\vec{r}_1 \, \frac{e^{i \vec{k}\cdot(\vec{r}-\vec{r}_1)}}{\vec{k}^2}\widetilde{\rho}(\vec{r}_1)\right] \, ,\nn \\
                      & = -G \beta A(R) \cos{(m \, t)} \int d^3 \vec{r}_1 \frac{\widetilde{\rho}(\vec{r}_1)}{| \vec{r} - \vec{r}_1|}\, , \nn \\
                      & = - \frac{G M}{r} \; \beta A(R) \; \cos{(m \, t )} \,,
\end{align}
the last line being valid only for $r\geq R_{source}$. 

Putting all results together, the total (zeroth- plus first-order) scalar field takes the form 
\begin{align}
\varphi(\vec{r},t) = A e^{- m^2 GM r}\left(1 - \frac{G M \beta }{r}\right)\cos{(m \, t )} \,.
\end{align}
This is clearly only true outside the matter distribution and on Solar System scales. In fact, in order to be consistent with the cosmological analysis of previous sections, we know that the amplitude of the zeroht-order scalar field does not vanish but instead it must asymptote to $\overline{\varphi}_0$ as $r \rightarrow \infty$. We thus match these two solutions at $r_c = \sqrt{GM/H_0}$, and find
\begin{align}
A = \overline{\varphi}_0 e^{m^2 GM \sqrt{\frac{GM}{H_0}}} \, ,
\end{align}
so that at first order
\begin{align}
\label{varphieq}
\varphi(\vec{r},t) = \overline{\varphi}_0 e^{m^2 GM \left(\sqrt{\frac{GM}{H_0}}-r\right)} \left( 1 - \frac{G M \beta}{r}\right)\cos{m \, t} \,.
\end{align}

This calculation is valid only for a certain range of scalar field masses $m$. In fact, our computation of the first order metric perturbations rests on the assumption that 
\begin{align}
m \varphi_{(0)} \thicksim \partial_0 \varphi_{(0)} \thicksim H_0 \, ,
\end{align}
which is to say that $\varphi_{(0)} \sim \overline{\varphi}_0 \cos{(m\,t )}$ and which is only true if the exponential in Eq.~\eqref{varphieq} is not too large. This in turn implies that
\begin{align}
m^2 GM \sqrt{\frac{GM}{H_0}} \lesssim 1 \, ,
\end{align}
or more exactly if this quantity is not large compared to unity. This means our calculations are valid provided 
\begin{align}
m \lesssim 10^{-15} \, \text{eV} \,.
\end{align}
One could extend the regime of validity of this solution, for example by going to next order in multiple scale analysis, but this is beyond the scope of this paper. 

In this mass domain and in the Solar System, one can therefore write
\begin{align}
\label{tot_field}
\varphi(\vec{r},t) = \overline{\varphi}_0 e^{m^2 GM \sqrt{\frac{GM}{H_0}}}\left[1-\frac{GM}{r}(\beta + m^2 r^2)\right] & + {\cal{O}}\left(\epsilon^2\right) \nn \\ & + {\cal{O}}\left(\overline{\varphi}_0^3 \epsilon\right) \,,
\end{align}
where the second order symbol, ${\cal{O}}\left(\overline{\varphi}_0^3 \epsilon\right)$, is emphasized to recall our previous assumption of neglecting terms that scale as $\overline{\varphi}_0^3$ times a first order quantity.

With the scalar field at hand, we can now solve for the metric tensor in the Jordan frame. From Eq.~\eqref{tot_field}, the Jordan frame metric takes the form
\begin{align}
d\widetilde{s}^2 & =  e^{\beta A^2 \cos{(mt)}^2}\left[1 - 2 \beta^2 A^2 \frac{GM}{r}\cos{(mt)}^2\right] ds^2\,, 
\end{align}
where we have neglected the $(m \, r)^2$ contribution appearing in Eq.~\eqref{tot_field} because this would give rise to a term of ${\cal{O}}(\varepsilon H_0^{2} R_S^2)$ -- an order we have neglected in this analysis. Since the overall conformal factor is closely related to the Jordan frame cosmological scale factor, we can also neglect it on astrophysical scales and we thus obtain
\begin{align}
& d\widetilde{s}^2  = -\left\{1 - \left[\chi - 2 \beta^2 A^2 \frac{G M}{r}\cos{(mt)}^2\right]\right\}dt^2 \nn \\ &  + \left\{1 + \left[\chi + 2 \beta^2 A^2 \frac{G M}{r}\cos{(mt)}^2\right]\right\}\delta_{ij}dx^idx^j + 2 \Gamma_i dx^idt \,.
\end{align}
We do not substitute Eq.~\eqref{eq:chi-approx} for $\chi$ in the above equation because Eq.~\eqref{eq:chi-approx} was obtained up to a correction factor of ${\cal{O}}(\overline{\varphi}_0^2)$, which although not important before, it must be taken into account now as we do below. Interestingly, and as expected from the intuition developed at the beginning of this section, the diagonal part of the metric takes roughly the same form as in the massless case, but with the substitution $\varphi_0 \rightarrow \overline{\varphi}_0 \cos{(mt)}$. 

%-----------------
\subsubsection{Constraints}
Let us now use the previous line element to see what constraints we can place on the parameters $m$ and $\beta$ of massive DEF theory using Solar System observations. We proceed in two steps. First, we compute the $\gamma_{\PPN}$ parameter of parametrized post-Newtonian theory. Second, we use assumptions of post-Newtonian theory to find the regime of validity of some of our expressions. The off-diagonal part of the metric is briefly discussed in Appendix ~\ref{app}, but it will not be used here. 

The diagonal part of the metric is characterized by $\chi$ which, from Eq.~\eqref{chi}, can be written at first order as
\begin{align}
\chi(r,t) = 2G \int d^3\vec{r}_1 \frac{\widetilde{\rho}(\vec{r}_1)}{|\vec{r}-\vec{r}_1|} e^{2 \beta A^2 \cos{[m(t - |\vec{r}-\vec{r}_1|)]}^2} \,.
\end{align}
This equation implies that structure dependent terms arise even at first post-Newtonian order in this theory, i.e., the solution to this integral explicitly depends on the radius of the considered body. For a test-particle with $\widetilde{\rho}(\vec{r})= M \delta(\vec{r})$, the above integral collapses to
\begin{align}
\chi(r,t) = \frac{2GM}{r}e^{2 \beta A^2 \cos{(m\, t)}^2} \,.
\end{align}
Therefore, the diagonal part of the metric gives rise to the following line element
\begin{align}
d\widetilde{s}^2_{diag}  &= - dt^{2} \left[1 - \frac{2GM}{r} e^{2 \beta A^2 \cos{(m\, t)}^2} 
\right. 
\nn \\ 
& \left. 
\qquad \qquad \left( 1 -  \beta^2 A^2\cos{(mt)}^2\right)\right]   
 \nn \\ &
 + \eta_{ij}dx^idx^j \left[1 + \frac{2GM}{r} e^{2 \beta A^2 \cos{(m\, t)}^2} 
\right. 
\nn \\ 
& \left. 
\qquad \qquad  
\left(1 + \beta^2 A^2 \cos{(mt)}^2\right)\right] \,.
\end{align}
up to $\mathcal{O}\left(A^4\right)$. Let us now make a few remarks about this line element. First, the effective gravitational constant $G_N$ appearing in the Solar System and measured by Cavendish-like experiments is
\begin{align}
G_N = G \, e^{2 \beta A^2 \cos{(m\, t)}^2} \left( 1 -  \beta^2 A^2\cos{(mt)}^2\right)\,,
\end{align}
which clearly is time-dependent. Furthermore, the effective $\gamma_{\PPN}$ parameter is
\begin{align}
\label{gammaPPN}
\gamma_{\PPN} = \frac{1 + \beta^2 A^2 \cos{(mt)}^2}{1 - \beta^2 A^2 \cos{(mt)}^2} \,.
\end{align}
Please note that, interestingly, on Solar System scales, all dependence in $r$ vanishes in $\gamma_{\rm PPN}$. This is very different from the results obtained when neglecting the cosmological evolution and the time dependence of the field in generic massive scalar-tensor theories as shown e.g. in~\cite{Hohmann:2013rba}. Both of these modifications allow, in principle, for constraints on the theory through observations of the Shapiro time delay from the Cassini spacecraft~\cite{lrr-2014-4}. 

First, when the mass of the scalar field is so small that the field evolves on time scales much larger than the observation time in the Solar System, and if today $m \, t \sim \pi/2$, then the theory reduces ``accidentally'' to GR. We do not consider this case further, as it requires fine-tuning; instead we make the assumption that $\cos{(mt)} = {\cal{O}}(1)$, which will lead to the strongest constraints. The rapid oscillations of the field [see e.g.~Eq.~\eqref{ODG}] suggests that this assumption is reasonable at least for $m\geq 10^{-20} \text{ eV}$. With this at hand, Eq.~\eqref{gammaPPN} and measurements of the Shapiro time delay require that
\begin{align}
\beta^2 A^2 \sim \beta^2 \overline{\varphi}_0^2 \lesssim 10^{-5} \, ,
\label{cassin}
\end{align}
so that
\begin{align}
\label{cons2}
%\beta^2 \frac{H_0^2}{m^2} \lesssim 10^{-5} \,,
\frac{m}{\beta} \gtrsim 10^{2.5} H_0 \sim 3 \times 10^{-31} \; {\textrm{eV}} \,,
\end{align}
is a \emph{sufficient} constraint for the theory to be in agreement with the measurements made by the Cassini spacecraft. Indeed we know that, assuming consistency of the cosmological evolution, the energy density of the scalar field can only be a fraction of the total measured energy density, i.e. 
\begin{align}
\label{density}
m^2 \overline{\varphi}_0^2 \lesssim H_0^2 \,,
\end{align}
and Eqs.~\eqref{cons2} and \eqref{density} imply Eq.~\eqref{cassin}. Clearly, this constraint is extremely weak, and thus, we conclude that massive DEF theory passes Solar System tests easily.

Furthermore, for the theory to be in agreement with the post-Newtonian expansion of GR, we need
\begin{align}
\frac{\partial_t h_{00}}{\partial_r h_{00}} \lesssim \mathcal{O}(\sqrt{\varepsilon}) \,.
\end{align}
In order of magnitude, the previous equation gives the requirement
\begin{align}
\beta (\beta - 1) (H_0 R_S) \frac{H_0}{m} \lesssim \sqrt{\varepsilon} \,.
\end{align}
For order of unity $\beta$ and in the limit where $m \gg H_0$, this constraints is automatically passed if the constraint given in Eq.~\eqref{cons2} is passed. One could also use the off-diagonal part of the metric to place a requirement on $m$ with Solar System observations, and we sketch this in Appendix~\ref{app}.

Our calculation suggests that, for $\beta \leq 0$ and with $|\beta|$ of order unity, adding a mass term to the Lagrangian of the scalar field allows for consistency with cosmological observations, which in turn implies consistency with Solar System experiments. Our analysis holds when $m \lesssim 10^{-15}$ eV and when the mass hierarchy $m \gg H_0 \sim 10^{-33} {\textrm{eV}}$ is satisfied. Consistency with Solar System tests imposes the very weak constraint $m/\beta \gtrsim 10^{-31} \text{ eV}$. The very weak constraint $m/\beta \gtrsim 10^{-31} \text{ eV}$ implies consistency with Solar System tests.

%%%%%%%%%%%%%%%%%%%%%%%%%%%%%%%%%%%
\section{Conclusions and Discussion}
In this paper, we studied massive DEF theory from the point of view of its cosmological evolution and its subsequent Solar System behavior. We have found that the scalar field generically goes through a quiescent phase at very early cosmological times, and then enters an exponentially-damped oscillatory phase, typically during the radiation-dominated era, that persists until today. Because of this, the cosmological value of the scalar field today, which serves as a boundary condition to astrophysical studies, is necessarily time-dependent. We have also calculated the behavior of the scalar field as sourced today by weak-gravitational sources to find that massive DEF theory is perfectly consistent with Solar System observations. Observations with the Cassini spacecraft do technically allow for a constraint on the mass of the scalar, but this constraint is extremely weak, leaving most of the parameter space available.  

Additionally, we have shown that the mass hierarchy $m \gg H_0$ leads to a consistent late-time cosmological scenario where the scalar field behaves as a (cold matter-like) pressureless fluid. In this context, the additional interactions caused by the scalar field could be ``hidden'' in the $\Omega_M$ term of cosmology, which describes the amount of baryonic and cold dark matter in a $\L CDM$ universe. One possible avenue for future work would be to extend the analysis presented here to more subtle cosmological effects, such as perturbations in the CMB and the growth of structure.

Perhaps one of the most important conclusions of our work is that in order to study astrophysical phenomena at small redshift, one must first carry out a careful matching between the cosmological and the astrophysical behavior of all fields. We have here presented a first attempt to understand the effect of the cosmological evolution of the scalar field on astrophysical scales, focusing only on Solar System observables. Another possible avenue for future work would be the study neutron star observables and spontaneous scalarization in neutron stars. Previous work in this area has assumed that the scalar field can be treated as static asymptotically at spatial infinity, but our work shows that for a wide range of scalar field masses, the scalar field remains time-dependent at late times. One could thus re-visit the study of spontaneous scalarization with a time-dependent ansatz for the scalar field that allows to naturally satisfy its time-dependent, cosmological boundary conditions. Similarly, one could also revisit the study of the super-radiant instability in massive scalar-tensor theories, which so far has only been carried out with time-independent boundary conditions.  These issues involve strong field analysis. As pointed out, the ansats we use may not be the correct one to study this regime. However, this paper provide a useful solution to compare against fully numerical studies of generic bodies in the weak field limit.

%%%%%%%%%%%%%%%%%%%%%%%%%%%%%%%%%%%
\section{Acknowledgements}
We would like to thank Enrico Barausse, Emanuele Berti, Hector Okada da Silva and David Anderson for many comments and suggestions that have greatly improved the quality and clarity of our paper. Our gratitude also goes to Gilles Esposito-Far\`ese from the IAP for his continuous mentoring to one of us. TAdPSA would like to thank the eXtreme Gravity Institute and the Department of Physics at Montana State University for their hospitality during the completion of this project. NY acknowledges support from the NSF CAREER grant PHY-1250636 and NASA grant NNX16AB98G. 

%%%%%%%%%%%%%%%%%%%%%%%%%%%%%%%%%%%
\appendix
\section{The Off-Diagonal Sector of the Metric}
\label{app}

In order to compute the $\Gamma_i$'s, let us require that $\Gamma_i(\vec{r},t)=f(r) (\partial r/\partial x^i) \sin{(2mt)}$. Recalling that $\partial r/\partial x^i$ is an eigenvector (with eigenvalue 2) of the Laplacian operator on the unit 2-sphere, Eq.~\eqref{gamma} implies
\begin{align}
4 m^2 f(r) + \frac{1}{r^2}\partial_r\left(r^2 \partial_r f\right) + \frac{2}{r^2}f(r) = 2 A^2 \frac{GM}{r^2}m \beta\,.
\end{align}
There is no simple analytical solution to this equation, but we can still perform an order-of-magnitude estimate of $f(r)$. When $m R_{\SSs} \ll 1$ (roughly when $m \ll 10^{-18}$ eV), the above evolution equation implies
\begin{align}
f(r) \sim (H_0 R_{SS}) \varepsilon A \beta\,.
\end{align}
The first order off-diagonal part of the metric therefore is a ``cosmological'' correction, and thus, it can be neglected on Solar System scales. On the other hand, when $m \, R_{\SSs} \gg 1$, one finds
\begin{align}
f(r) \sim \varepsilon \beta \frac{A^2}{m R_{\SSs}}\,.
\end{align}
In order to be consistent with the post-Newtonian expansion of GR in the Solar System, however, we must require that
\begin{align}
\frac{A^2}{m R_{\SSs}} \lesssim \mathcal{O}(\sqrt{\epsilon})\,,
\end{align}
which leads to the requirement
\begin{align}
\frac{H_0^2}{m^2}\frac{1}{m R_{\SSs}} \lesssim 10^{-4}
\end{align}
This inequality is almost redundant with the requirement that $m R_{\SSs} \gg 1$. This implies that we can neglect off-diagonal corrections of the metric at first order and in the range of mass parameter we are considering. 

%%%%%%%%%%%%%%%%%%%%%%%%%%%%%%%%%%%%%%%%%%%%
%%%%%%%%%%%%%%%%%%%%%%%%%%%%%%%%%%%%%%%%%%%%
\bibliography{master_bis}
\end{document}